\documentclass[12pt,preprint,apj]{emulateapj}

\newcommand{\bfr}{f_{bar}}
\newcommand{\ha}{H\alpha}
\newcommand{\brho}{\bar{\rho}}
\newcommand{\rtwenty}{\rho_{20}}
\newcommand{\distnei}{R_{n}/r_{vir,n}}
\newcommand{\bfrsbo}{f_{\rm SB1}}
\newcommand{\bfrsbw}{f_{\rm SB2}}
\newcommand{\ana}{A\&A}

\shorttitle{Bar Fraction of Galaxies}
\shortauthors{Lee et al.}

\begin{document}

\title{Dependence of Barred Galaxy Fraction on Galaxy Properties and Environment}

\author{Gwang-Ho Lee}
\affil{Department of Physics and Astronomy, Seoul National University, Gwanak-gu, Seoul 151-742, Korea}
\email{ghlee@astro.snu.ac.kr}

\author{Changbom Park}
\affil{Korea Institute for Advanced Study, Dongdaemun-gu, Seoul 130-722, Korea}
\email{cbp@kias.re.kr}

\author{Myung Gyoon Lee}
\affil{Department of Physics and Astronomy, Seoul National University, Gwanak-gu, Seoul 151-742, Korea}
\email{mglee@astro.snu.ac.kr}
\and

\author{Yun-Young Choi}
\affil{Department of Astronomy \& Space Science, Kyung Hee University, Kyungki 446-701, Korea}
\email{yy.choi@khu.ac.kr}

%==============================================================================================================

\begin{abstract}

We investigate the dependence of occurrence of bars in galaxies on galaxy properties and environment.
We use a volume-limited sample of 33,391 galaxies brighter than $M_{r}=-19.5+5$log$h$ at $0.02\le z\le0.05489$,
drawn from the SDSS DR 7.
We classify the galaxies into early and late types, and identify bars by visual inspection.
Among 10,674 late-type galaxies with axis ratio $b/a>0.60$, we find 3,240 barred galaxies ($\bfr=30.4\%$)
which divide into 2,542 strong bars ($\bfrsbo=23.8\%$) and 698 weak bars ($\bfrsbw=6.5\%$).
We find that $\bfrsbo$ increases as $u-r$ color becomes redder, and that it has a maximum value at
intermediate velocity dispersion ($\sigma\simeq$150 km s${}^{-1}$). This trend suggests that strong bars are
dominantly hosted by intermediate-mass systems.
Weak bars prefer bluer galaxies with lower mass and lower concentration. In the case of
strong bars, their dependence on the concentration index appears only for massive galaxies with $\sigma>150$ km s${}^{-1}$.
We also find that $\bfr$ does not directly depend on the large-scale background density when other physical parameters
($u-r$ color or $\sigma$) are fixed. We discover that $\bfrsbo$ decreases as the separation
to the nearest neighbor galaxy becomes smaller than 0.1 times the virial radius of the neighbor regardless of neighbor's
morphology. These results imply that strong bars are likely to be destroyed during strong tidal interactions, and that
the mechanism for this phenomenon is gravitational and not hydrodynamical. The fraction of weak bars has no correlation
with environmental parameters.
We do not find any direct evidence for environmental stimulation of bar formation.

\end{abstract}


\section{Introduction}

Stellar bars are common among spirals, and are believed to have an important role in the evolution
of their host galaxies. Several studies presented ideas that bars channel gas from the outer disk into the nuclear region
\citep{huntley78,knapen95,hunt99,saka99,jogee99,jogee05,sheth05,elme09}, and that they redistribute angular momentum
of the baryonic and dark matter components of disk galaxies \citep{lyndenbell79,sellwood81,albada81,combes85,
weinberg85,dns00,ath2003,agu2009}.
Bars are also thought to have a significant role in fueling of active galactic nuclei \citep{schlos89,ho97}
and in forming bulges or pseudo-bulges \citep[e.g.,][]{combes81,pfenniger84,knk2004,sheth05,martinez+06,deb+05,deb+06,mend+08,agu2009}.

Some simulations showed that bars can be destroyed by a large central mass concentration
\citep{roberts79,norman96,snm99,ath2005}. This result indicates two possibilities.
First, it is possible that currently non-barred spirals had a bar in the past \citep{knk2004}.
Second, bars may be recurrent \citep{bournaud2002,berentzen2004,gad06}. However, there are conflicting results
that bars are dynamically robust structures, requireing large mass concentrations to dissolve bars
\citep{dns00,shen2004,deb+06}.
Therefore it remains an open question why some spirals have a bar structure, while others do not.

Most previous studies tried to explain the formation and evolution of bars through internal secular evolution
only. To get a better understanding of bars, we need to consider not only internal but also external influence
on their evolution. \citet{thompson81} showed that the fraction of barred galaxies is significantly larger
in the core of the Coma cluster than in the outer part of the cluster, and \citet{bnv90} argued that the strong tidal field
in the core of Coma can transform non-barred spirals into barred ones. Some studies claimed through N-body
simulations that bars are triggered by external effects such as tidal interactions \citep{gca90} and the passage
of a companion galaxy \citep{berentzen2004}. Then it is expected that the bar fraction ($\bfr$)
may be different in different environments.

There have been observational efforts to find the relation between environment and the bar fraction. \citet{esk00}
found that the bar fraction in the Fornax and Virgo clusters is slightly higher than the average value for fields.
\citet{vdb02} also suggested from the analysis of 930 galaxies in the northern Shapley-Ames catalog that $\bfr$
in cluster environments is larger than in groups or fields. The sizes of their sample were not large enough to
derive a statistically meaningful result, so that they concluded that dependence of $\bfr$ on environment seems to be
unclear. Recently some studies used large samples including thousands of galaxies drawn from
the Sloan Digital Sky Survey (hereafter, SDSS; \citealt{york00}), but environmental dependence of the bar
fraction is still controversial \citep{agu2009,cheng09,giordano10,mabreu10,cameron10,barway10}.

Recently, there have been efforts to find relations between bars and diverse properties of their host galaxies
from statistical approaches using large galaxy samples. \citet{barazza08} presented that $\bfr$ in $r$-band is
relatively higher in later type spirals that are bluer and less-concentrated systems.
\citet{agu2009} also obtained a similar result. However, \citet{masters10,masters11} found
an opposite result that the optical bar fraction significantly increases as ($g-r$) color becomes redder and they suggested
a color bimodality in barred galaxies. Furthermore, \citet{nair10b} found that the bar fraction shows a bimodal
distribution of stellar mass with a break at log($M/M_{\odot}$)$\sim10.2$, and also presented that the bar fraction
can be described as a function of the concentration of galaxies.

The purpose of this paper is to examine how the existence of bar structure is related with the physical
properties of galaxies and to find what kind of environmental conditions the barred galaxies prefer. We use
a sample of galaxies in the SDSS, and measure the bar fraction as a function of internal properties of galaxies such
as luminosity, color, star formation rate, concentration, and central velocity dispersion. The environmental
parameters considered include the local galaxy mass density, distance to the nearest neighbor galaxy, and the morphology of the neighbor galaxy.

This paper is composed as follows. Section 2 gives a brief description of the volume-limited sample used in this study,
and introduces physical and environmental parameters of galaxies. Morphological classification of barred galaxies and
comparison with previous classifications are described in $\S$3. In $\S$4, we present results of the dependence of $\bfr$
on physical and environmental parameters of galaxies. In $\S$5, we discuss the implication of our
results, and $\S$6 summarizes our primary results.

\section{Data}

\subsection{A Volume-limited Sample}

\begin{figure}
\figurenum{1}
\centering
\includegraphics[scale=0.55]{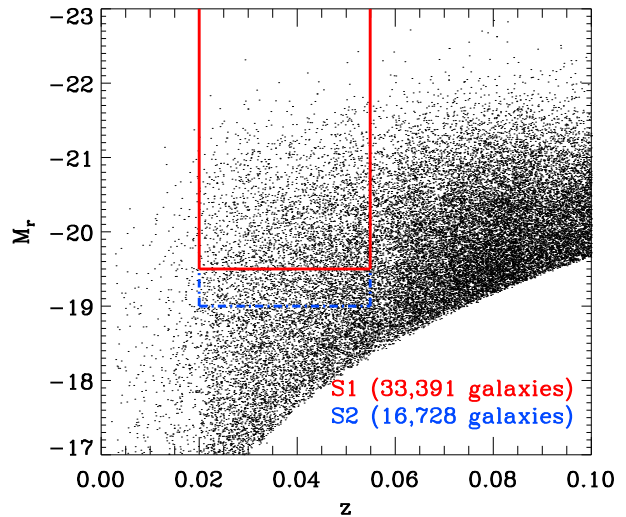}
\caption{Definition of our volume-limited SDSS samples in redshift versus $r$-band absolute
magnitude space. The S1 sample is defined by $M_{r}<-19.5$ and $0.02\le z\le 0.05489$, and
the S2 sample is defined by $-19.0>M_{r}\ge -19.5$ and $0.02\le z\le 0.05489$. The S1 sample and the S2
sample consist of 33,391 galaxies and 16,728 ones, respectively.}
\label{vol_limited}
\end{figure}

%% Table 1
\begin{deluxetable}{ccccc}
\tablecolumns{5}
\tablewidth{0pc}
\tablecaption{Morphological Classes of the Sample Galaxies}
\tablehead{
\colhead{Sample} & \colhead{Total} & \colhead{Early-type} & \colhead{Late-type} & \colhead{Unclassified}}
\startdata
S1 & 33,391 & 13,867 & 19,431 & 93 \\
S2 & 16,728 & 5,334 & 11,393 & 1 \enddata
\label{table_morph}
\end{deluxetable}

\begin{figure}
\figurenum{2}
\centering
\includegraphics[scale=0.45]{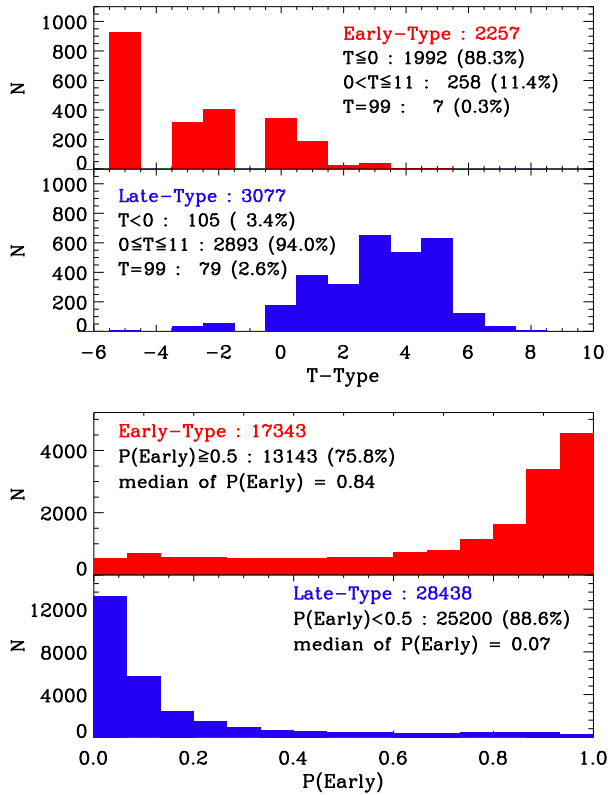}
\caption{A comparison of early- and late-type classification with two previous works: ({\it upper}) T-Types in \citet{nair10a}
and ({\it lower}) $P({\rm Early})$ in \citet{huertas+11}.}
\label{compmc}
\end{figure}

We use a volume-limited sample of 50,119 galaxies brighter than $M_r=-19.0+5{\rm log}h$ mag (hereafter,
we drop the $+5{\rm log}h$ term in the absolute magnitude) at redshift $0.02 \leq z \leq 0.05489$, drawn
from the SDSS Data Release 7 (DR7; \citealp{abaza09}). These galaxies are extracted from the Korea Institute
for Advanced Study Value-Added Galaxy Catalog (KIAS VAGC; \citealp{choi+10}), which is a catalog based on the
Large Scale Structure (LSS) sample of New York University Value-Added Galaxy Catalog (NYU VAGC; \citealp{blan05}).
Our volume-limited sample is based on the D2 sample introduced in \citet{choi+07}.
The D2 sample contains galaxies brighter than $M_{r}=-18.5+5{\rm log}h$ at $0.025<z<0.05485$,
and drawn from the SDSS DR5 \citep{adelman+07}.

We divide our volume-limited sample into two subsamples according to the galaxy magnitudes as shown
in Figure \ref{vol_limited}: a bright S1 sample containing galaxies with $M_{r}<-19.5$ mag,
and a faint S2 sample including galaxies with $-19.0>M_{r}\ge -19.5$ mag. The S1 is the major subsample
in our study, and the S2 is an auxiliary subsample used when studying the environmental effects.

First, we classify all galaxies in our sample into early types (E/S0) and late types (S/Irr)
using the automated method introduced by \citet{pnc05}. They found that the $u-r$ color versus
$g-i$ color gradient and the $u-r$ color versus concentration index spaces can be used
to classify galaxies into early- and late-type galaxies.
The reliability and completeness of this automated method is about 90\%.
Morphology information given by this scheme has been used in many studies on relation between galaxy properties
and the environment \citep{choi+07,choi+09,park07,park08,pnc09,pnh08,lee+08,lee+10a,lee+10b,lee+10c,han+10,cervantes10,cervantes11,
hwang+10,hwang+11,hwangnpark10,lara+10,tortora+10}.
To improve the accuracy of morphology classification we perform additional visual check.
The morphology of 2,427 galaxies in the S1 sample, which is about 7.3\%,
is changed by the additional visual check. As a result, the S1 sample includes 13,867 early types, 19,431
late types, and 93 unclassified galaxies, while the S2 sample consists of 5,334 early types,
11,393 late types, and 1 unclassified galaxy, as listed in Table \ref{table_morph}. We exclude
unclassified galaxies in the following analysis.

We compare our classification with two previous works: \citet[][hereafter NA10]{nair10a} and \citet{huertas+11}.
We find a good agreement among three morphology classifications. Figure \ref{compmc} shows the results.

NA10 presented a visual classification for $\sim14,000$ galaxies in the SDSS DR4 with $g<16$ mag
at $0.01<z<0.1$. They assigned T-Types to all galaxies in their sample: T-Type$=0$ for S0/a,
$<0$ for E and S0, $1-9$ for Sa-Sm, $10-11$ for Im, and 99 for unknown galaxies (peculiar galaxies, or unclassified ones).
For 5,334 galaxies included in both our and NA10's sample, we classify these galaxies into 2,257 early-type galaxies
and 3,077 late-type galaxies. The majority (88.3\%) of these early-type galaxies are assigned to T-Type$\leq0$ (E or S0),
while 11.4\% of them have T-Types later than S0/a and 0.3\% of them are classified as unknown galaxies.
On the other hand, 94\% of the late-type galaxies have T-Types for S0/a-Sm, and only 105 galaxies (3.4\%) have T-Types
earlier than S0/a, and 79 galaxies (2.6\%) are classified as unknown galaxies.

\citet{huertas+11} presented a Bayesian automated classification for $\sim700,000$ galaxies from the SDSS DR7
spectroscopic sample, and they quantified probabilities to each galaxy of being in four types (E, S0, Sab, Scd).
Probabilities of being in early-type galaxy, $P({\rm Early})$, are only used in this comparison.
For 45,781 galaxies common in our and their sample, we divide them into 17,343 early-type and
28,438 late-type galaxies. We find that median values of $P({\rm Early})$ are 0.84 and 0.07 for the early-type
and the late-type galaxies, respectively. In addition, 75.8\% of early-type galaxies have $P({\rm Early})$ higher than 0.5,
while 88.6\% of late-type galaxies have $P({\rm Early})$ lower than 0.5.
In conclusion, our morphology classification shows a good agreement with two previous works.

\subsection{Physical Parameters of Galaxies}

We use the physical parameters of galaxies to study the dependence of $\bfr$ on the properties of galaxies:
absolute Petrosian magnitude ${}^{0.1}M_r$, ${}^{0.1}(u-r)$ color, Petrosian radius ($R_{Pet}$) in $i$-band,
color gradient in ${}^{0.1}(g-i)$, inverse concentration index ($c_{in}$), central velocity dispersion ($\sigma$),
and equivalent width of the $\ha$ line. These parameters reflect
most of major physical properties of galaxies from morphology and mass to kinematics and star formation
activity \citep{pnc09}.

The rest-frame absolute magnitudes of individual
galaxies are computed in fixed bandpass, shifted to $z=0.1$, using the Galactic reddening correction of
\citet{schle1998} and K-corrections as described by \citet{blan03}. The mean evolution correction given
by \citet{teg04}, $E(z)=1.6(z-0.1)$, is also applied.
The superscript 0.1 indicates that the absolute magnitude is the rest-frame magnitude when the galaxy
is at $z=0.1$, after K-correction and luminosity evolution correction.
To compute colors, we use extinction and K-corrected model magnitudes.
${}^{0.1}(u-r)$ and ${}^{0.1}(g-r)$ colors also follow this convention. Hereafter, the superscript 0.1 will be omitted.
The $g-i$ color gradient is the difference between the color with aperture radius $R<0.5R_{Pet}$ and the color
with the annulus $0.5R_{Pet}<R<R_{Pet}$. When the color gradient has a negative value, it indicates that
the galaxy has a redder inner part and a bluer outer part. The inverse concentration index ($c_{in}$) is defined
by $R_{50}/R_{90}$ where $R_{50}$ and $R_{90}$ are semi-major axis lengths of ellipses enclosing $50\%$
and $90\%$ of Petrosian flux in the $i$-band image, respectively. The central velocity dispersion value
is adopted from NYU-VAGC \citep{blan05}. The value of $\ha$ equivalent width is taken from MPA/JHU-VAGC
\citep{trem04}, which was computed using the straight integration over the fixed bandpass from the
continuum-subtracted emission line with the model of \citet{bruz03}.
We adopt a flat ${\rm \Lambda CDM}$ cosmology with $\Omega_{\Lambda}=0.74$ and $\Omega_{m}=0.26$
which are from WMAP 5-year data \citep{komatsu2009}.

\subsection{Environmental Parameters}

We use three environmental parameters. One is the mass density measured by using twenty neighboring galaxies.
This is called the large-scale background density, or briefly background density.
The second is the distance to the nearest neighbor galaxy normalized by the virial radius of the neighbor.
The third is the morphology of the nearest neighbor galaxy.

\subsubsection{Large-Scale Background Density}

The background density at a given location of a galaxy is measured by
\begin{equation}
\rtwenty(x)/\brho=\sum_{i=1}^{20}\gamma_{i}L_{i}W_{i}(|x_{i}-x|)/\bar\rho,
\end{equation}
where $x$ is the location of the galaxy, and $\gamma_i$, $L_i$, and $x_i$ are mass-to-light ratio, luminosity,
and position of the closest twenty galaxies brighter than $M_{r}=-19.0$ mag (S1+S2, see Figure
\ref{vol_limited}). Note that our study will focus on galaxies brighter than $M_{r}=-19.5$ mag (S1).
The mass associated with the galaxy plus dark halo system is assumed to be proportional to the luminosity
of the galaxy. The mean mass density is obtained by
\begin{equation}
\brho=\sum_{all}\gamma_{i}L_{i}/V,
\end{equation}
where $V$ is the survey volume. \citet{pnc09} obtained $\brho=0.02255(\gamma L)_{-20}(h^{-1}$Mpc)$^{-3}$,
where $(\gamma L)_{-20}$ is the mass of a late-type galaxy with $M_{r}=-20$ mag.
We assume $\gamma$ (early-type) $=2\gamma$ (late-type) at the same $r$-band luminosity.
We do not need to know the absolute value of $\gamma$'s since $\gamma$ appears in both numerator and denominator
in Equation (1). We use the spline-kernel weight $W$ described in detail in \citet{park07} for the background
density estimation. We vary the size of the spline kernel to include twenty galaxies within the kernel.

\subsubsection{The Nearest Neighbor}
To find the effects of the nearest neighbor galaxy on bar formation we use the distance
to the nearest neighbor galaxy and its morphology.
For a target galaxy with absolute magnitude $M_r$ in the S1 sample, the nearest neighbor galaxy is the one
that has the smallest projected separation from the target galaxy among galaxies brighter than $M_{r}+0.5$ mag
and with a radial velocity difference smaller than $V_{max}$. We adopt $V_{max}=600$ and 400 km s${}^{-1}$
for the early- and late-type target galaxies, respectively \citep{park08}.
Note again that the target galaxies have $M_{r}<-19.5$ mag and their neighbors have $M_{r}<-19.0$ mag.

The small-scale density environment experienced by a target galaxy attributed to its nearest neighbor
is estimated by
\begin{equation}
\rho_{n}/\bar{\rho}=\gamma_{n}L_{n}/(4\pi R_{n}^{3}\bar{\rho}/3),
\end{equation}
where $R_n$ is the projected separation of the neighbor from the target galaxy,
$\gamma_n$ and $L_n$ are the mass-to-light ratio and the $r$-band luminosity of the neighbor, respectively.
We define the virial radius of a galaxy as the radius where the mean mass density within a sphere with that
radius is 200 times the critical density or 740 times the mean density of the universe \citep{pnc09}, namely,
\begin{equation}
r_{vir}=(3\gamma L/4\pi/740\bar{\rho})^{1/3}h^{-1}Mpc.
\end{equation}
In Section 4 we will use the parameter, $R_{n}/r_{vir,n}$, the nearest neighbor distance normalized
by the virial radius of the neighbor.

\begin{figure*}
\figurenum{3(A)}
\centering
\includegraphics[scale=1.0]{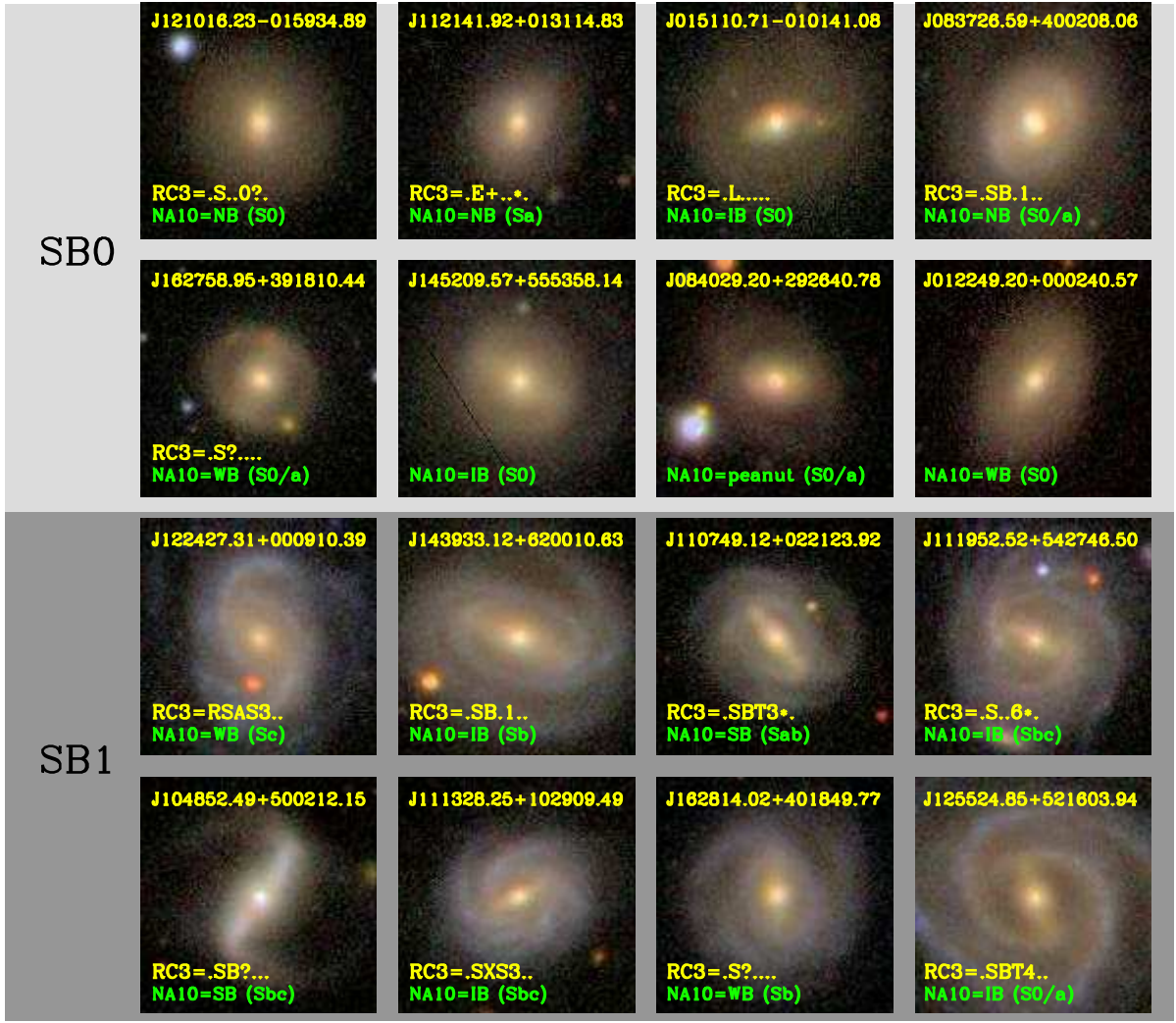}
\caption{Color images of example galaxies belonging to four classes: SB0 ({\it upper}), SB1 ({\it lower}), SB2, and SB3.
The RC3 classification type or NA10's classification type are shown at the bottom of each image whenever available.
NA10 classified barred galaxies into several types : strong barred (SB), intermediate barred (IB), weak barred (WB),
and non-barred (NB). Notations given in the parentheses represent the morphology of host galaxies.}
\label{exam_bar1}
\end{figure*}

\begin{figure*}
\figurenum{3(B)}
\centering
\includegraphics[scale=1.0]{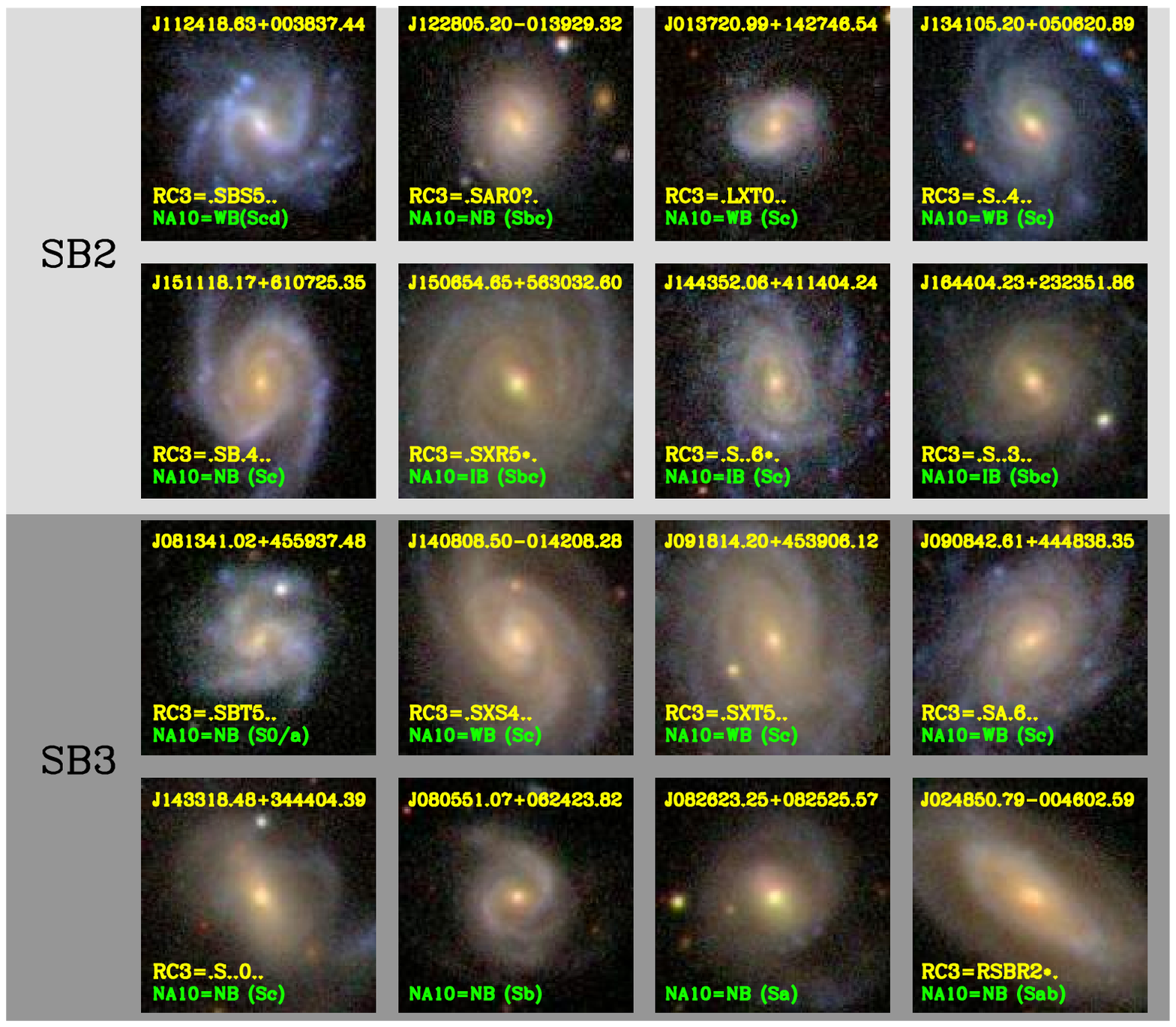}
\caption{Same as Figure \ref{exam_bar1}, but for SB2 ({\it upper}) and SB3 ({\it lower}).}
\label{exam_bar2}
\end{figure*}

\begin{figure}
\figurenum{4}
\centering
\includegraphics[scale=0.7]{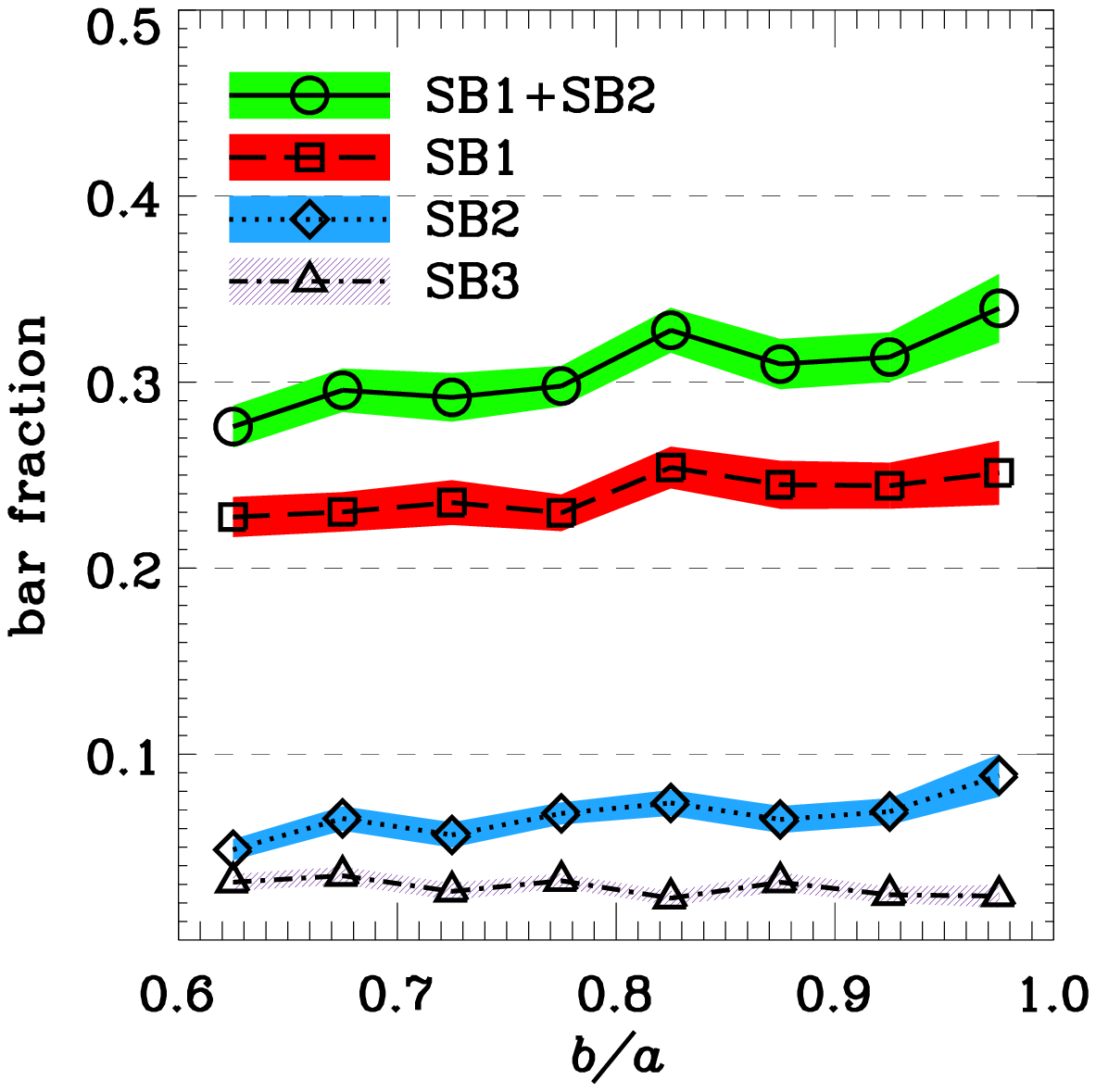}
\caption{Axis ratio dependence of the bar fraction. Squares, diamonds, triangles represent the bar fraction in an
axis ratio bin for SB1, SB2, and SB3 galaxies, respectively, while circles are sum for SB1 and SB2 galaxies.
Shades mean 1-$\sigma$ sampling errors estimated by calculating the standard deviation of the bar fraction
in 1,000-times-repetitive sampling.}
\label{abtrue_bf}
\end{figure}

\begin{figure}
\figurenum{5}
\centering
\includegraphics[scale=0.6]{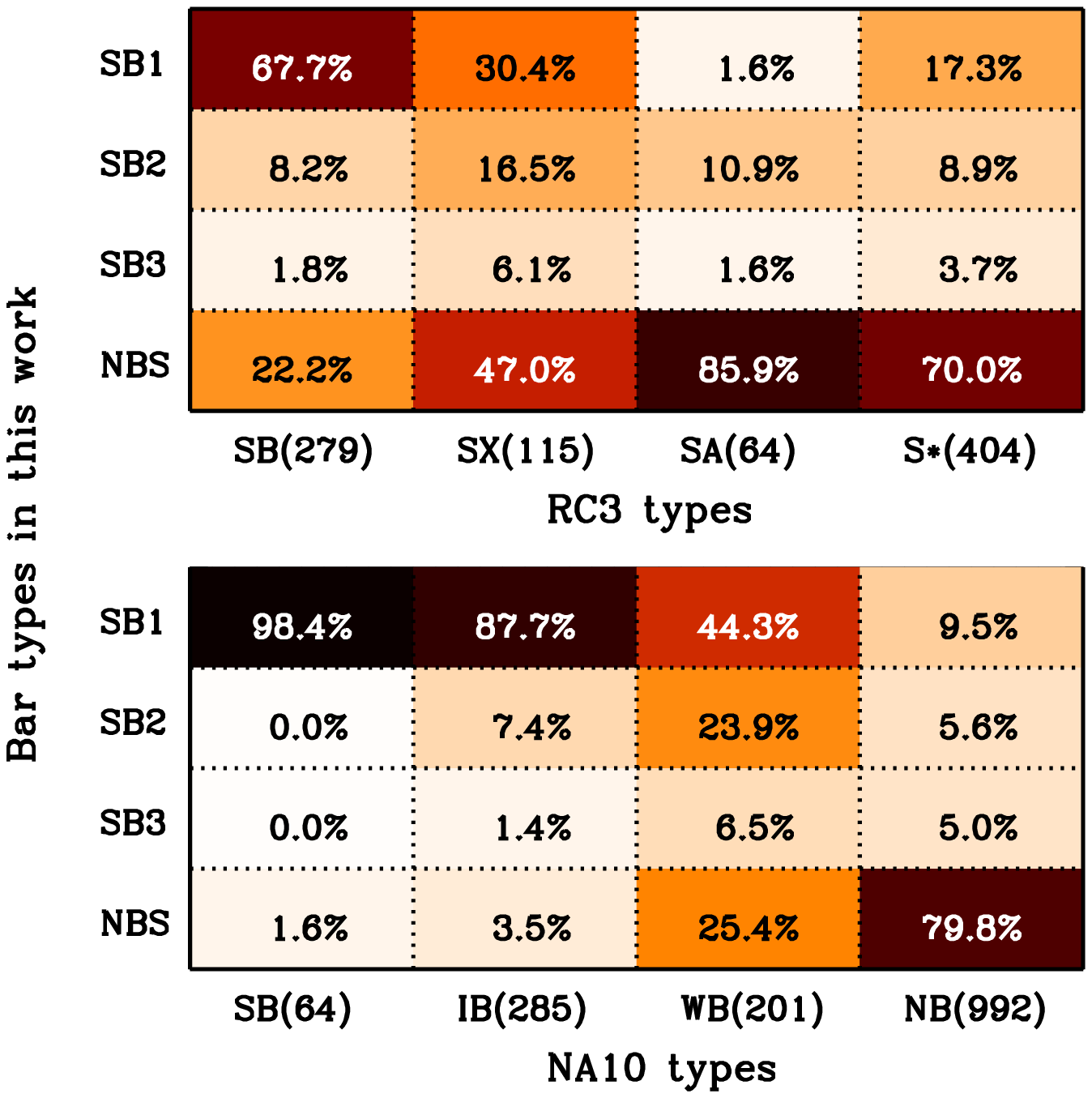}
\caption{Lists of percentage of each morphological type in our classification (SB1$\sim$SB3 \& NBS:
non-barred spirals) for a subclass of late-type galaxies classified by RC3 ({\it upper panel}) and NA10 ({\it lower panel}).}
\label{comp_rc3_nair}
\end{figure}

\section{Selection of Barred Galaxies}

Various methods have been used to find barred galaxies in previous studies.
Visual inspection is a traditional and classical method used for long (\citealp{kna99,esk00,vdb02,sheth08,mabreu10,masters10,masters11};
NA10, \citealp{nair10b}).
The Revised Shapley-Ames Catalog of Bright Galaxies \citep{snt81} provides the ``gold standard'' of
galaxy classification based on visual inspection of blue light images. The Reference Catalog of Bright
Galaxies (de Vaucouleurs et al. 1991, hereafter RC3) also classified morphological types of galaxies
using visual inspection. Using these galaxy catalogs, some studies investigated the differences between
barred and non-barred galaxies \citep{kna99,esk00,vdb02}.

Recently, bar structures are identified and characterized by new techniques such as the ellipse \citep{jed87} fitting
method \citep{wozniak95,jogee99,jogee+02,jogee+04,knapen+00,del07,maj07,ree07,sheth08,barazza08,agu2009,cheng09}, and Fourier analysis
\citep{lau04,lau06,lau07,agu2009}.
These automated methods can be applied to large samples of galaxies, and can determine the properties
of bars quantitatively.

There is a significant difference in the bar fraction depending on the method for selecting bars.
The ellipse fitting method suggests that the bar fraction is nearly 50\% in optical bands :
$\bfr=44\%\pm7\%$ in B-band \citep{maj07}, 47\% in I-band \citep{ree07}, $\sim48\%-52\%$ \citep{barazza08}
and $45\%$ \citep{agu2009} in $r$-band. However, the optical bar fraction obtained by visual inspection is
significantly lower: $\bfr=25.3\%$ \citep{snt81} and $\sim33\%$ (RC3) in B-band,
$26\%\pm0.5\%$ (NA10) and $29.4\%\pm0.5\%$ \citep{masters10} in SDSS $g+r+i$ combined color
images. \citet{masters11} pointed out that this discrepancy is caused by differences
in wavelengths and selection criterion.
It is possible that the ellipse fitting method classify many ovals or central
distorted features as bars. However, it is not obvious what causes such a difference at the moment.
Thus future works are needed to investigate differences between visual inspection
and the ellipse fitting method.

\subsection{Morphological Classes of Barred Galaxies}

We select barred galaxies by visual inspection of $g+r+i$ combined color images (hereafter color images).
Color images are obtained from {\it http://cas.sdss.org} using {\it Visual Tools}.
We use the color images because the stellar population within a bar is generally different from
that in other part of its host galaxy and because it is easier to visually identify a bar
when color information is available.

The selection of barred galaxies is performed on both early-type and late-type galaxies.
For late-type galaxies, we adopt an $i$-band isophotal axis ratio limit, $b/a>0.6$, to reduce the internal extinction
effects and the selection bias due to the inclination. Thus, the number of selected late-type galaxies in the
S1 sample is 10,674.

We classify visually selected barred galaxies into four classes based on the bar strength taking into
account the structural properties of host galaxies. In the case of late-type barred galaxies
we estimate relative sizes of bars to their host galaxies by eyes, and divide them into strong-barred (SB1),
weak-barred (SB2), and ambiguous barred galaxies (SB3).
The criteria of classification and the number of galaxies in each class are as follows.

(1) SB0 : 905 early-type galaxies having a bar structure and no spiral arm. Therefore, they are barred
lenticular galaxies.

(2) SB1 : 2,542 late-type galaxies having a strong bar. The size of bars is larger than one quarter of
the size of their host galaxies. Generally these galaxies are early-type or intermediate-type spirals
with well-developed arms and a relatively large bulge.

(3) SB2 : 698 late-type galaxies having a weak bar. The size of bars is smaller than one quarter of the size
of their host galaxies. Host galaxies are typically late-type spirals with a small bulge.

(4) SB3 : 401 late-type galaxies for which it is difficult to decide whether it is barred or not.

Figures \ref{exam_bar1} and \ref{exam_bar2} display color images of some typical galaxies belonging to
the four classes. We also show RC3 classification type or NA10 classification type whenever available.

Figure \ref{abtrue_bf} shows the fraction of galaxies in each bar class as a function of $b/a$ axis ratio.
Highly inclined galaxies have small $b/a$, while nearly face-on galaxies have a value close to 1.
When $b/a>0.6$, $\bfr$ remains nearly constant in each bar class. This means that our morphology classification of
face-on late-type galaxies with $b/a>0.6$ provides a reliable and robust $\bfr$ that is unaffected by the inclination
of galaxies.

\subsection{Comparison with Previous Works}

We compare the results of our classification with those of two previous works: the morphology classifications
in RC3 and NA10. RC3 classified barred galaxies into three classes : strong barred (SB), weak barred (SX),
and non-barred (SA). However, some galaxies have no information of a bar: they are described just as `S.'
or `S?'. So, we assign `S.' or `S?' galaxies to one group `S*' as shown in Figure \ref{comp_rc3_nair}.
Upper panel of Figure \ref{comp_rc3_nair} lists the percentage of each morphological type in our classification
for each subclass of late-type galaxies classified in the RC3 catalog. For example, there are 279 SB-type
galaxies common in RC3 and our sample. Among them we classify 67.7\% (189 galaxies) as SB1, 8.2\% (23 galaxies)
as SB2, 1.8\% (5 galaxies) as SB3, and 22.2\% (62 galaxies) as non-barred systems.
There are total 862 galaxies used in the whole comparison. It shows that correspondence between RC3 and our classification
is not very high. Among 279 SB-type galaxies defined by RC3, 62 galaxies (22.2\%) are classified as non-barred systems in
our classification. Also, we classify only 46.9\% of 115 SX-type galaxies as the systems having an obvious bar.

Lower panel of Figure \ref{comp_rc3_nair} lists the percentage of each morphological type in our classification
for a subclass of late-type galaxies defined by NA10. NA10 classified barred galaxies into four types : strong barred (SB),
intermediate barred (IB), weak barred (WB), and non-barred (NB).
For 1,542 late-type galaxies that are common in our sample and NA10's catalog, it shows a good agreement
between our and NA10's morphological classifications.
Especially it shows above 95\% correspondence for SB-type and IB-type galaxies. For 201 WB-type galaxies we classify 137 galaxies
(68.2\%) as SB1 or SB2 classes, while we classify 13 (6.5\%) and 51 galaxies (25.4\%) as SB3 and non-barred galaxies, respectively.
When considering the SB3 galaxies as non-barred systems, the agreement rate between NA10's and our classification is about 85\%.
Note that, in this comparison, $\bfr$ is 36.0\% in NA10 classification, while 40.3\% (for SB1+SB2) in our classification.
Therefore, it is concluded our study and NA10 agree well in morphological classifications and bar fractions.

\subsection{The Fraction of Barred Galaxies}

%%Table 2
\begin{deluxetable}{ccccc}
\tablecolumns{5}
\tablewidth{0pc}
\tablecaption{Fraction of Barred Galaxies}
\tablehead{
\colhead{Galaxy type} & \colhead{$N_{total}$} & \colhead{Bar class}& \colhead{$N_{bar}$} & \colhead{$\bfr$}}
\startdata
Early types & 13,867 & SB0 & 905 & 6.5\% \\
\\
Late types & 10,674 & SB1 & 2,542 & 23.8\% \\
($b/a>0.6$) & & SB2 & 698 & 6.5\% \\
 && SB3 & 401 & 3.8\%  \enddata
\label{table_bf_vis}
\end{deluxetable}

\begin{figure}
\figurenum{6}
\centering
\includegraphics[scale=0.6]{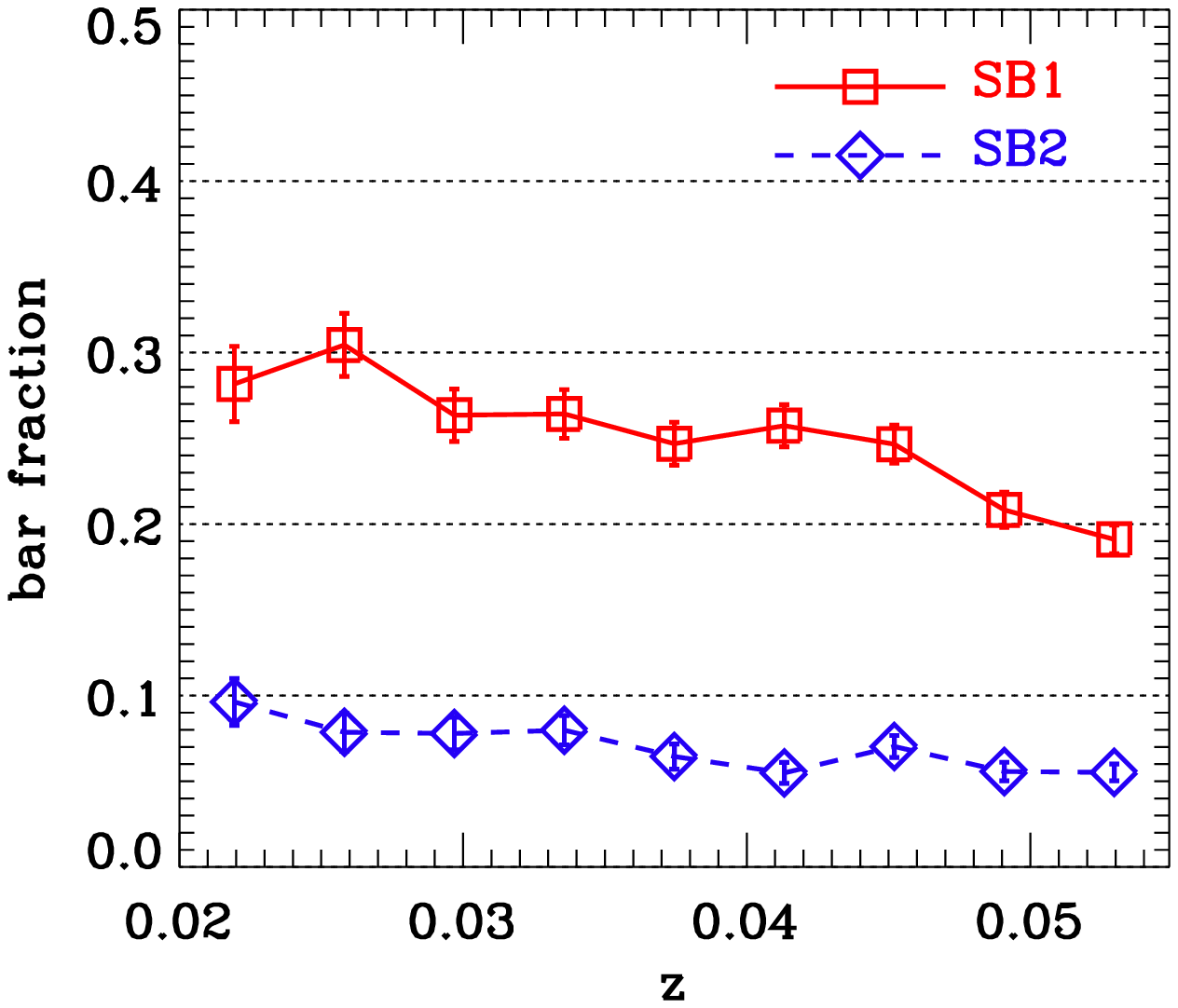}
\caption{Dependence of the bar fraction on redshift. Squares and diamonds represent the bar fraction for SB1 and SB2 galaxies, respectively.
Error bars represent 1-$\sigma$ of 1,000-times-repetitve sampling.}
\label{zdepend}
\end{figure}

\begin{figure*}
\figurenum{7}
\centering
\includegraphics[scale=0.6]{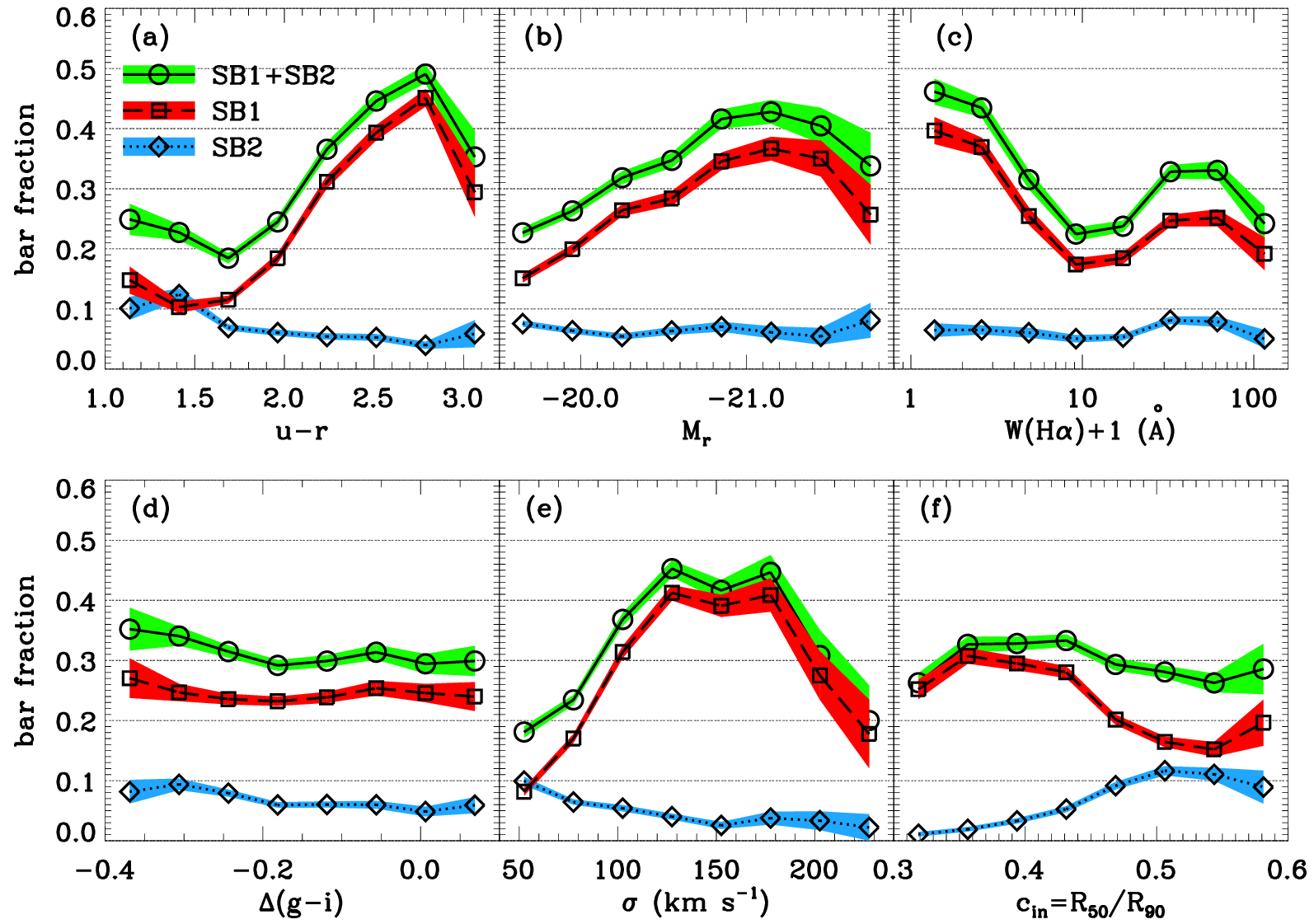}
\caption{Dependence of the bar fraction for late-type galaxies on (a) $u-r$ color, (b) $r$-band absolute magnitude,
(c) equivalent width of the $\ha$ line, (d) color gradient, (e) central velocity dispersion, and (f) inverse concentration index.
Squares and diamonds represent, respectively, fraction of SB1 (strong bar) and SB2 (weak bar) galaxies.
Circles represent the sum of both types. Shades mean 1-$\sigma$ sampling errors.}
\label{params_bf}
\end{figure*}

Among 13,867 early-type galaxies, there are 905 SB0 galaxies (6.5\%), while we find 3,641 barred galaxies belonging to
SB1-SB3 types among the 10,674 late-type galaxies with $b/a>0.60$ ($\bfr=34.1\%$) as shown in Table \ref{table_bf_vis}.
In some studies, $\bfr$ is defined as the the frequency of barred galaxies among disk galaxies including both spirals and
lenticulars \citep{esk00,maj07,ree07,barazza08}. In this study, however, we do not distinguish lenticulars from early-type
galaxies. Therefore, hereafter, we define $\bfr$ as the percentage of barred galaxies among late-type galaxies
with $b/a>0.60$. We do not use 905 SB0 galaxies when calculating the bar fraction of late type galaxies.

Among the three types for late-type barred galaxies (SB1-SB3), we regard only two types, SB1 and SB2, as barred galaxies.
SB3 galaxies have an elongated feature in their central region, but it is uncertain to consider it as a bar.
Some SB3 galaxies seem to have oval structures in their center. In previous studies some oval galaxies were classified
as barred galaxies, but generally they are considered as non-barred ones \citep{knk2004}.
We consider SB3 types as non-barred galaxies from now on.
In conclusion, we find that the bar fraction for late-type galaxies is 30.4\% (3,240 barred galaxies).
This value is in good agreement with recent studies that used visual inspection to select barred galaxies, 25$\sim$33\%
(NA10; \citealp{giordano10,masters11}).

Figure \ref{zdepend} shows the dependence of the bar fraction on redshift.
We find that the fraction of SB1 galaxies ($\bfrsbo$) is approximately constant, $\sim25\%$,
until $z=0.045$, but that it decreases to $\sim20\%$ at $z>0.045$.
On the other hand, the fraction of SB2 galaxies ($\bfrsbw$) vary insignificantly from 9\% to 6\%
as the redshift increases. However, even if we restrict our analysis
to 5,888 galaxies at $z<0.045$, we find that there is no significant change in the
following results.

\section{Results}

\begin{figure*}
\figurenum{8}
\centering
\includegraphics[scale=0.6]{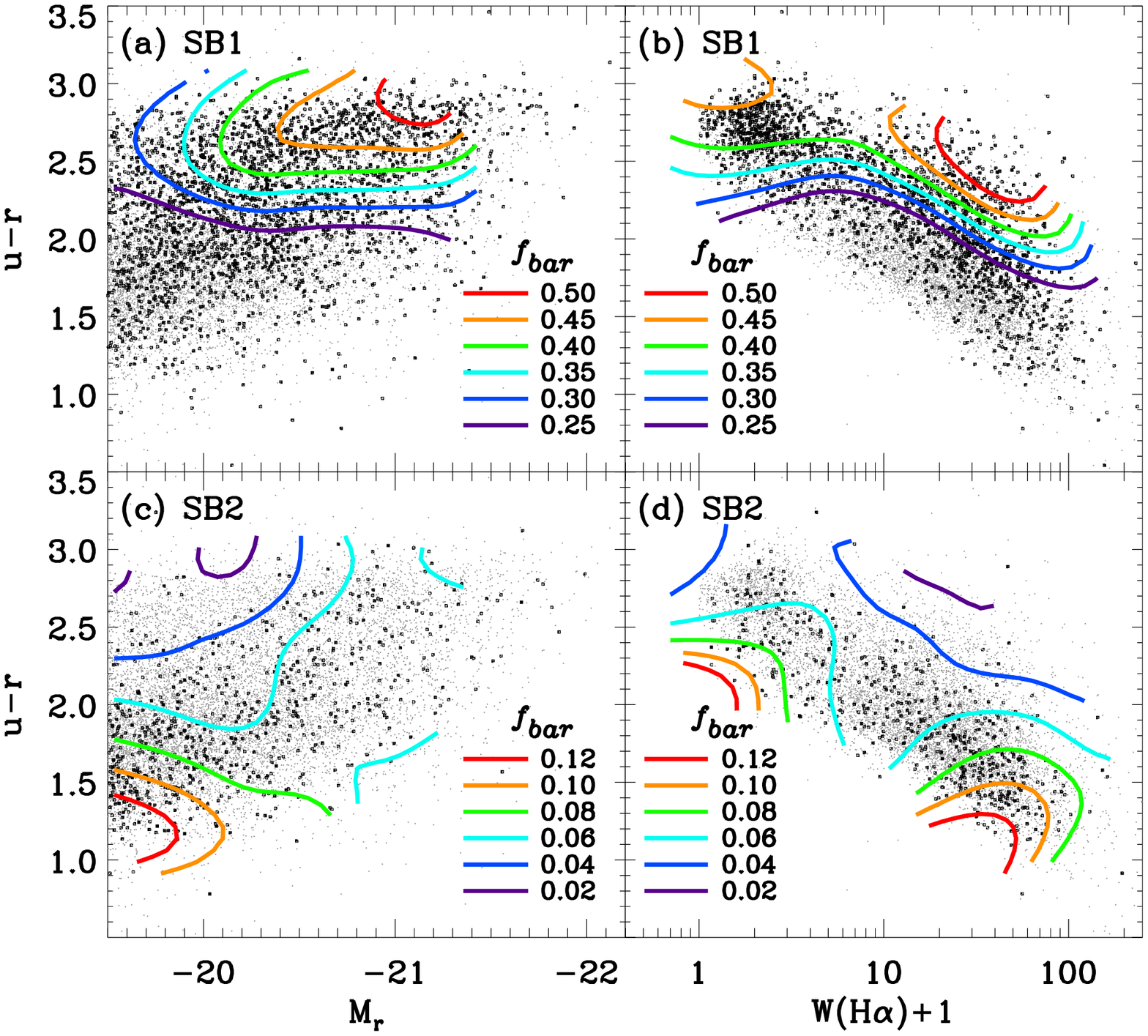}
\caption{The bar fraction ($\bfr$) contours in ({\it left}) the $u-r$ color versus the $r$-band absolute magnitude $M_r$ space and
in ({\it right}) the $u-r$ color versus the equivalent width of H$\alpha$ line space. Black dots in upper panels and lower panels
represent SB1 galaxies and SB2 galaxies, respectively. Contours in panel (a) and (b) represent the SB1 galaxy fraction,
while the SB2 galaxy fraction in panel (c) and (d).}
\label{ur_rabsmag_lineHa}
\end{figure*}

\begin{figure*}
\figurenum{9}
\centering
\includegraphics[scale=0.6]{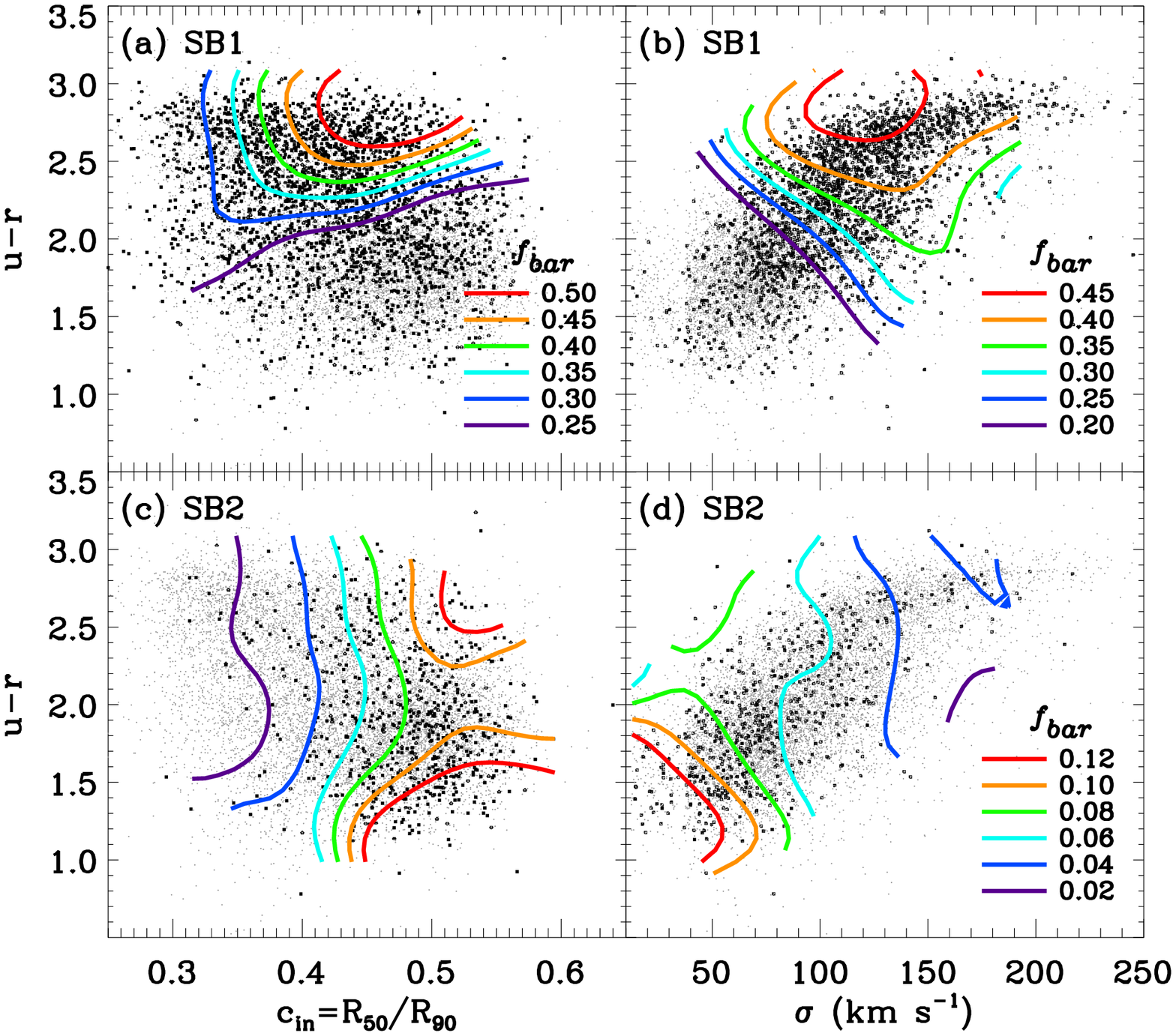}
\caption{The bar fraction ($\bfr$) contours in ({\it left}) the $u-r$ color versus the ({\it inverse}) concentration index $c_{in}$
space and in ({\it right}) the $u-r$ color versus the central velocity dispersion ($\sigma$) space. Contours in panel (a) and
(b) show the SB1 galaxy fraction, while the SB2 galaxy fraction in panel (c) and (d). Contour levels in panel (c) represent
the same as those in panel (d).}
\label{ur_conx_sigma}
\end{figure*}

\begin{figure*}
\figurenum{10}
\centering
\includegraphics[scale=0.6]{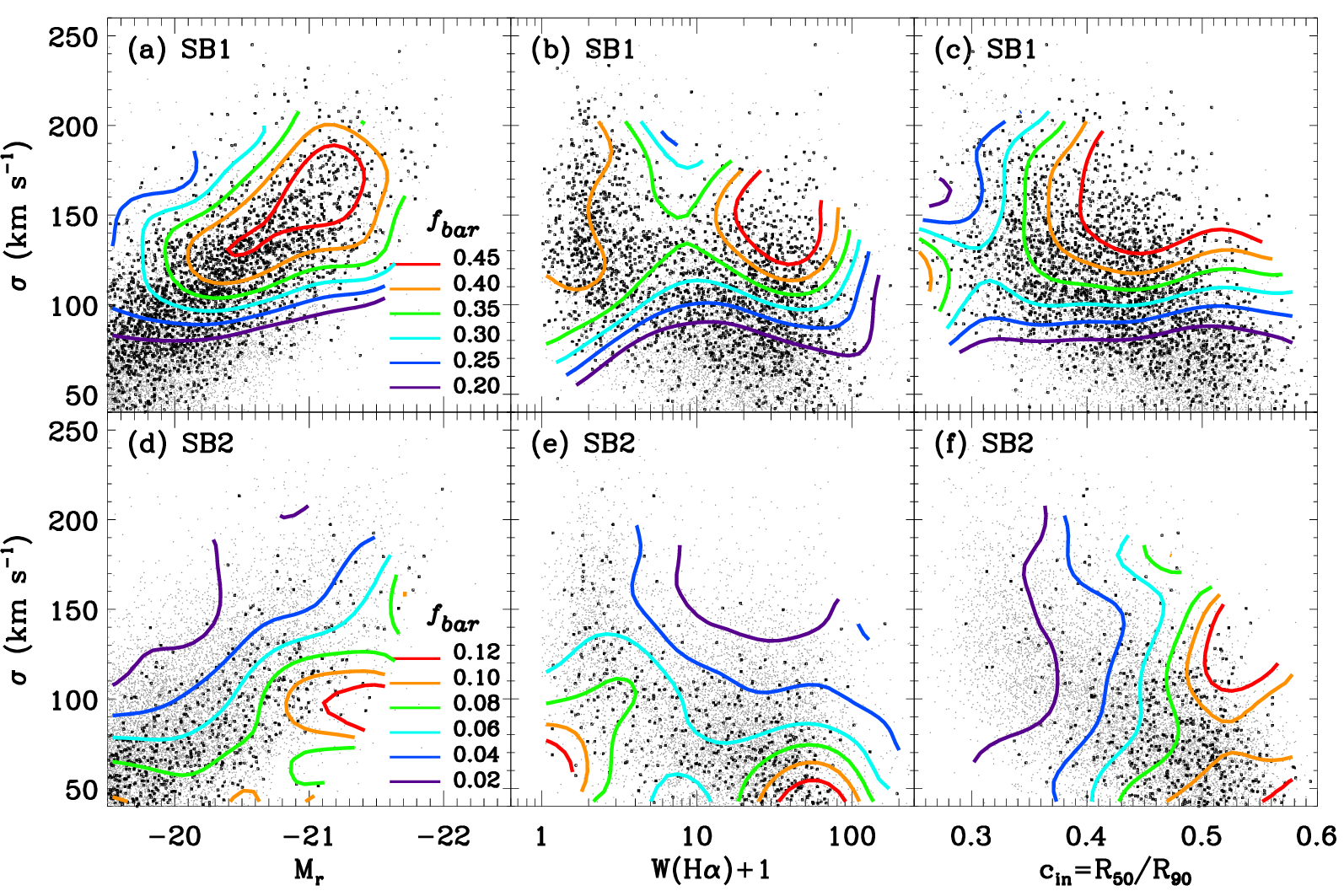}
\caption{Fraction of SB1 ({\it upper}) and SB2 ({\it lower}) galaxies (contours) in two-dimensional parameter spaces:
(a) the central velocity dispersion $\sigma$ versus the $r$-band absolute magnitude $M_r$, (b) $\sigma$ versus the $\ha$
equivalent width, and (c) $\sigma$ versus the (inverse) concentration index $c_{in}$.}
\label{params_vdisp}
\end{figure*}

\subsection{Dependence of the barred galaxy fraction on galaxy properties}

We investigate the dependence of $\bfr$ on several physical parameters of galaxies:
$r$-band absolute magnitude, $u-r$ color, $g-i$ color gradient, equivalent width of the $H\alpha$ line ($W(H\alpha)$),
$i$-band inverse concentration index ($c_{in}=R_{50}/R_{90}$), and central velocity dispersion ($\sigma$).
Figure \ref{params_bf} presents how the fractions of SB1 and SB2 galaxies change with variation of several
physical parameters. In the case of spectroscopic parameters such as $W(H\alpha)$ and the central velocity
dispersion, we measure $\bfr$ using data satisfying signal-to-noise ratio $S/N\ge10$. We estimate the errors
of $\bfr$ by calculating the standard deviation in 1,000-times-repetitive sampling method.

Figure \ref{params_bf}a shows $\bfr$ as a function of $u-r$ color.
The fraction of SB1 galaxies increases significantly as $u-r$ increases.
The SB1 fraction is less than $20\%$ for galaxies bluer than $u-r=2.0$, but it reaches $45\%$ when $u-r\simeq2.8$.
Then it decreases at the red color end. This result implies that passively evolving red late-type galaxies are more likely
to have a strong bar than blue spirals with some star formation activity. In contrast, the fraction of SB2 galaxies ($\bfrsbw$)
appears to become the maximum when $u-r\simeq1.4$. Then, $\bfrsbw$ becomes smaller as $u-r$ increases. These trends are consistent
with the result of \citet{hoyle+11} who found that longer bars than 5$h^{-1}$ kpc prefer redder late-type galaxies, and that
shorter bars inhabit bluer late-type galaxies.

In Figure \ref{params_bf}b, we plot $\bfr$ as a function of $r$-band absolute magnitude.
It is seen that $\bfrsbo$ monotonically increases as $M_r$ decreases until it reaches the maximum
at $M_{r}\simeq-21.2$ mag, and then slightly drops at the brightest magnitudes. On the other hand, the SB2 fraction shows no
dependence on $M_r$. \citet{mabreu10} also found a similar result using $\sim$190 galaxies in the Coma cluster.

In Figure \ref{params_bf}c, we plot the relation between $\bfr$ and $H\alpha$ equivalent width.
$W(H\alpha)$ is an indicator of star formation rate (SFR) in the central region ($R<1^{\prime\prime}.5$) of galaxies.
There were some reports that bars induce central starburst \citep{hunt99,esk00,jogee05,maj07}.
Therefore, it is expected that $\bfr$ is higher for galaxies having large $W(H\alpha)$.
However, our result is not exactly same as the expectation. We find two distinguishable components with peaks at
$\bfrsbo$ at $W(\ha)\simeq50 {\rm \AA}$ and at $W(\ha)\simeq0$ \AA. This indicates the existence of two populations:
SB1 galaxies with higher SFR or with very low SFR. In contrast, in the case of SB2 galaxies, a slight $\bfrsbw$-excess is
seen at $W(\ha)\sim50$ \AA. In general $\bfrsbw$ shows little dependence on $W(H\alpha)$.

Figure \ref{params_bf}d shows the dependence of $\bfr$ on color gradient $\Delta (g-i)$.
If a galaxy has a red inner part and blue outer part, $\Delta (g-i)$ is negative.
Bar structures are believed to be dominated by older stellar population relative to the outer disk region
of galaxies \citep{esk00}. Because we focus on late-type galaxies, it is reasonable that $\Delta (g-i)$ has
a range from $\sim$0 to --0.4 \citep{pnc05}. Both $\bfrsbo$ and $\bfrsbw$ shows no strong dependence on $\Delta (g-i)$.

Figure \ref{params_bf}e shows the dependence of $\bfr$ on central velocity dispersion $\sigma$.
At first $\bfrsbo$ increases as $\sigma$ increases. It reaches a peak of $\sim$40\% at intermediate central velocity
dispersion ($\sigma\simeq150\pm25$ km s$^{-1}$), and then it drops significantly when $\sigma$ is larger than $\sim175$ km
s$^{-1}$. This result suggests that strong bars are dominantly hosted by galaxies having intermediate central velocity
dispersion and that bars could be destroyed or not have formed when the central parts of galaxies are supported by stars
with the velocity dispersion exceeding $\sim$175 km s$^{-1}$.
The central velocity dispersion is closely associated with the mass of galaxies.
Therefore, this result is consistent with previous studies suggesting that more bars are detected in
intermediate mass galaxies \citep{sheth08,cameron10,mabreu10}. In contrast $\bfrsbw$ is a decreasing function of $\sigma$. It is 10\%
at $\sigma=50$ km s$^{-1}$, but nearly zero when $\sigma>150$ km s$^{-1}$. It means that weak bars mainly inhabit low mass systems.

Figure \ref{params_bf}f shows the relation between $\bfr$ and inverse concentration index $c_{in}=R_{50}/R_{90}$.
The concentration index is closely related with the Hubble sequence \citep{anm00,agu2009}.
A majority of early-type (E/S0) galaxies have a typical value of $\sim0.3$, while late-type (S/Irr) galaxies
cover a broad range of $c_{in}$ from 0.3 to 0.6 (see Fig.1 of \citealt{pnc05}).
As $c_{in}$ decreases, $\bfrsbo$ becomes larger. However, $\bfrsbw$ shows an opposite trend.
Generally galaxies having a smaller $c_{in}$ show earlier type morphologies compared to galaxies with a larger $c_{in}$.
Therefore, this result indicates that the bars within early- or intermediate-type spirals in the Hubble sequence
are larger than those hosted by late-type spirals. This result was also found in previous studies \citep{del07,agu2009}.

In Figure \ref{params_bf} we find that the fraction of SB1 and SB2 can be described as functions of six
physical parameters. The SB1 fraction strongly depends on $u-r$, $M_r$, $\ha$, and $\sigma$. Also, it shows relatively weaker
dependence on $c_{in}$. But, there is no dependence on $\Delta(g-i)$.
Whereas $\bfrsbw$ varies depending on the concentration index, there is no or weak correlations with other parameters.
In fact, variation of a physical parameter involves variation of another parameter. The $u-r$ color is indicative of recent star
formation so that it is correlated with the equivalent width of the $H\alpha$ line.
In addition, brighter galaxies have higher velocity dispersion \citep{park07}. Some physical parameters,
such as color, color gradient, and concentration vary along the Hubble sequence simultaneously \citep{pnc05}.
Therefore, we need to investigate the distribution of barred galaxies in multi-dimensional parameter spaces
to distinguish which parameters affect the variation of $\bfr$ directly or indirectly.

First, we check the distribution of barred (SB1 and SB2) and non-barred late-type galaxies
in multi-parameter spaces involving the $u-r$ color and other physical properties in Figure \ref{ur_rabsmag_lineHa}
and \ref{ur_conx_sigma}. We use a spline kernel to obtain a smooth distribution of $\bfr$ in each two dimensional
parameter space as shown by contours.

In Figure \ref{ur_rabsmag_lineHa}a, we find that $\bfrsbo$ depends on both $u-r$ color and $r$-band absolute magnitude.
The SB1 fraction is nearly zero for the bluest and faintest galaxies, and it increases monotonically
from lower left corner to upper right corner. When $u-r\le2.5$, color dependence is dominant.
The SB1 fraction increases as $u-r$ increases at any fixed $M_r$. Its dependence on $M_r$ is weaker than on color.
Stronger $M_r$-dependence of $\bfrsbo$ appears at only $u-r>2.5$: $\bfrsbo$ rises from 25\% to 50\% as $M_r$ decreases.
In contrast, SB2 galaxies are bluer than SB1 galaxies as shown in Figure \ref{ur_rabsmag_lineHa}c.
The fraction of SB2 galaxies has a dependence on $u-r$ color, but, its $M_r$-dependence is uncertain.
The color dependence of $\bfrsbw$ is obvious at $M_r>-20.5$, but this trend becomes unclear at $M_r<-20.5$.

Figure \ref{ur_rabsmag_lineHa}b and \ref{ur_rabsmag_lineHa}d show how $\bfrsbo$ and $\bfrsbw$ vary in $u-r$ color
and $W(H\alpha)$ space.
First, in the case of $\bfrsbo$, we find that strong dependence on the $u-r$ color appears to be dominated.
At any fixed $W(H\alpha)$, $\bfrsbo$ always increases as $u-r$ color becomes redder.
At $W(H\alpha)<9{\rm \AA}$, contours are nearly horizontal, and there is no dependence on $W(H\alpha)$.
However, at $W(H\alpha)\ge9{\rm \AA}$, contours are skewed from upper left to lower right.
At a given $u-r$ color, $\bfrsbo$ increases as $W(H\alpha)$ increases from $\sim10{\rm \AA}$ to $\sim100{\rm \AA}$.
This result is similar to the trend shown in Figure \ref{params_bf}c that there is an increase of the SB1 fraction
at $W(\ha)>9{\rm \AA}$. This result is consistent with previous prediction that bars may play a role in driving gas
towards the center and in triggering starburst in the central regions \citep{quillen95,saka99,esk00,jogee05,sheth05}.
On the other hand, $\bfrsbw$ increases as $u-r$ color becomes bluer as shown in Figure \ref{ur_rabsmag_lineHa}d.
This color dependence appears clearly at $W(\ha)\sim0{\rm \AA}$ or at $W(\ha)>20{\rm \AA}$. However, $W(\ha)$-dependence
is not obvious unlike the case of $\bfrsbo$ at even higher $W(\ha)$.

Figure \ref{ur_conx_sigma} displays $u-r$ color versus the concentration index ($c_{in}$) and $u-r$ color versus the central
velocity dispersion ($\sigma$) distribution for barred (SB1, SB2) and non-barred late-type galaxies.

We find that $\bfrsbo$ depends on both $u-r$ color and $c_{in}$. When $c_{in}>0.43$ the color dependence is dominant: $\bfrsbo$
increases as $u-r$ also increases. For highly concentrated populations with $c_{in}<0.43$, the color dependence almost disappears
and $\bfrsbo$ varies depending on only the concentration index. It is noted that $\bfrsbo$ decreases significantly
as $c_{in}$ decreases, when $u-r>2.0$. Interestingly it is not consistent with the result shown in Figure \ref{params_bf}f.
It indicates that the trend of $\bfrsbo$ seen in Figure \ref{params_bf}f is mainly generated by a large contribution of
blue galaxies with $u-r<2.0$. On the other hand, SB2 galaxies are mainly located at lower right side in Figure \ref{ur_conx_sigma}c. At $c_{in}\lesssim0.48$, contours are nearly vertical, which means that $c_{in}$-dependence of $\bfrsbw$ is much stronger than
$u-r$ color-dependence. The SB2 fraction also increases as $c_{in}$ increases like $\bfrsbo$ at $u-r>2.0$.
We find that the bar fraction decreases in highly concentrated galaxies, regardless of the strength of bars.

Figure \ref{ur_conx_sigma}b and \ref{ur_conx_sigma}d display $\bfrsbo$ and $\bfrsbw$ in $u-r$ versus $\sigma$ spaces.
The fraction of SB1 galaxies is again an increasing function of $u-r$ color. In addition, particularly,
the $\sigma$-dependence of $\bfrsbo$ is also shown. When $u-r$ color is fixed, $\bfrsbo$ increases until it reaches a peak
at $\sigma\simeq125$ km s$^{-1}$, and it decreases as $\sigma$ increases when $\sigma>150$ km s$^{-1}$.
Such tendency is already found in Figure \ref{params_bf}e. The central velocity dispersion is related with
the mass of their host galaxies. Therefore, at a fixed color, galaxies having intermediate mass have
a relatively higher possibility to host a strong bar.
While the fraction of SB2 galaxies shows negative correlations with both $u-r$ color and the velocity dispersion.
In this panel, it becomes obvious that weak bars prefer low-mass systems with small $\sigma$ and that their host galaxies
have generally bluer colors.

We find in Figures \ref{ur_rabsmag_lineHa} and \ref{ur_conx_sigma} that both $\bfrsbo$ and $\bfrsbw$ have
a strong dependence on $u-r$ color. Dependence on $u-r$ color is always seen in all panels involving four other parameters.
It is very likely that color is an independent parameter determining the bar fraction of late-type galaxies.
We also find that the central velocity dispersion is another important parameter in determining $\bfr$.
Therefore, it is needed to study the relation between the bar fraction and $\sigma$ in more detail.

Figure \ref{params_vdisp} shows the $\sigma$-dependence of $\bfrsbo$ and $\bfrsbw$. In the upper three panels, we find the strong
dependence of $\bfrsbo$ on $\sigma$: $\bfrsbo$ increases as $\sigma$ increases when $\sigma\lesssim125$ km s$^{-1}$.
Figure \ref{params_vdisp}a shows that $\bfrsbo$ reaches a peak at $\sigma\simeq175$ km s$^{-1}$. The locus of peak tends to shift
from $\sigma\simeq$ 125 km s${}^{-1}$ to $\sim$175 km s$^{-1}$ as galaxies become brighter. Figure \ref{params_vdisp}b shows
that there are two regions where $\bfrsbo$ has a maximum value at $W(H\alpha)=0$ ${\rm \AA}$ and $W(\ha)\simeq30$ \AA.
In Figure \ref{params_vdisp}c, we find a strong dependence on $c_{in}$ at $\sigma>150$ km s$^{-1}$.
For massive galaxies with $\sigma\geq$150 km s$^{-1}$, $\bfrsbo$ becomes smaller in highly concentrated systems than in less
concentrated ones. On the other hand, $\bfrsbw$ increases as $\sigma$ decreases as shown in Figure
\ref{params_vdisp}d and e. However, in Figure \ref{params_vdisp}f, we find a strong $c_{in}$-dependence of $\bfrsbw$.

In summary, we find three influential parameters on the bar fraction: $u-r$ color, central velocity dispersion, and
concentration index. The SB1 fraction increases as $u-r$ color becomes redder, and it has a maximum value at intermediate
$\sigma$. In contrast, the SB2 fraction becomes larger in galaxies with bluer color and lower $\sigma$. On the other hand,
higher concentration (smaller $c_{in}$) reduces both $\bfrsbo$ and $\bfrsbw$. The $c_{in}$-dependence of $\bfrsbo$ appears
only for red and massive galaxies with $u-r>2.0$ and $\sigma>150$ km s${}^{-1}$.

\subsection{Dependence of the barred galaxy fraction on environmental parameters}

\begin{figure*}
\figurenum{11}
\centering
\includegraphics[scale=0.7]{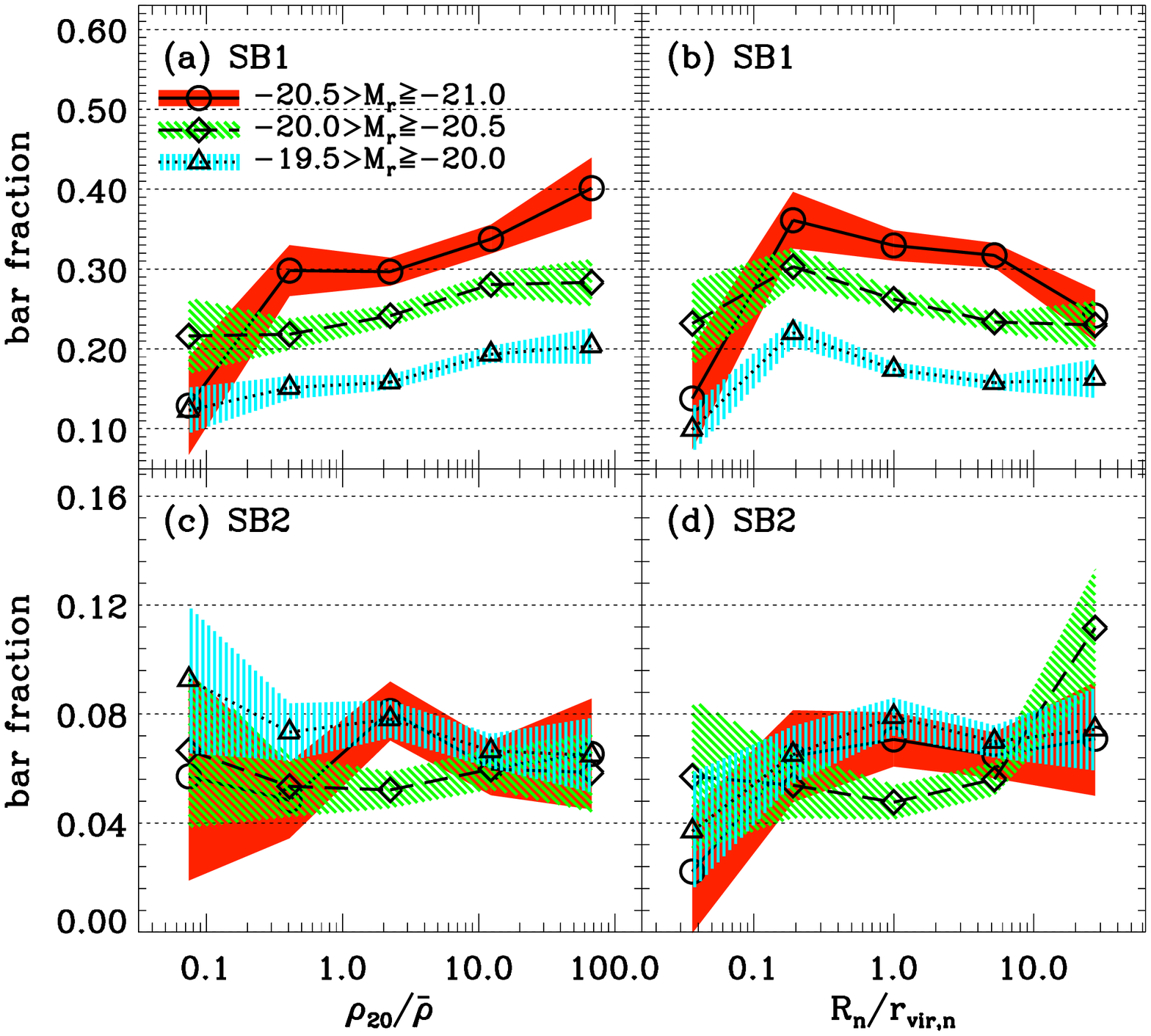}
\caption{Fraction of barred galaxies among face-on ($b/a>0.60$) late-type barred galaxies as a function of (a) the
large-scale background galaxy density ($\rtwenty/\brho$) and (b) the distance to the nearest neighbor normalized by
the nearest neighbor's virial radius ($\distnei$) in fixed absolute magnitude ranges. The absolute magnitude bins are
$-19.5>M_{r}\ge-20.0$ (triangles connected by a dotted line), $-20.0>M_{r}\ge-20.5$ (diamonds connected by a dashed line),
and $-20.5>M_{r}\ge-21.0$ (circles connected by a solid line). Shades represent 1-$\sigma$ sampling errors.}
\label{envir_bf}
\end{figure*}

\begin{figure*}
\figurenum{12}
\centering
\includegraphics[scale=0.7]{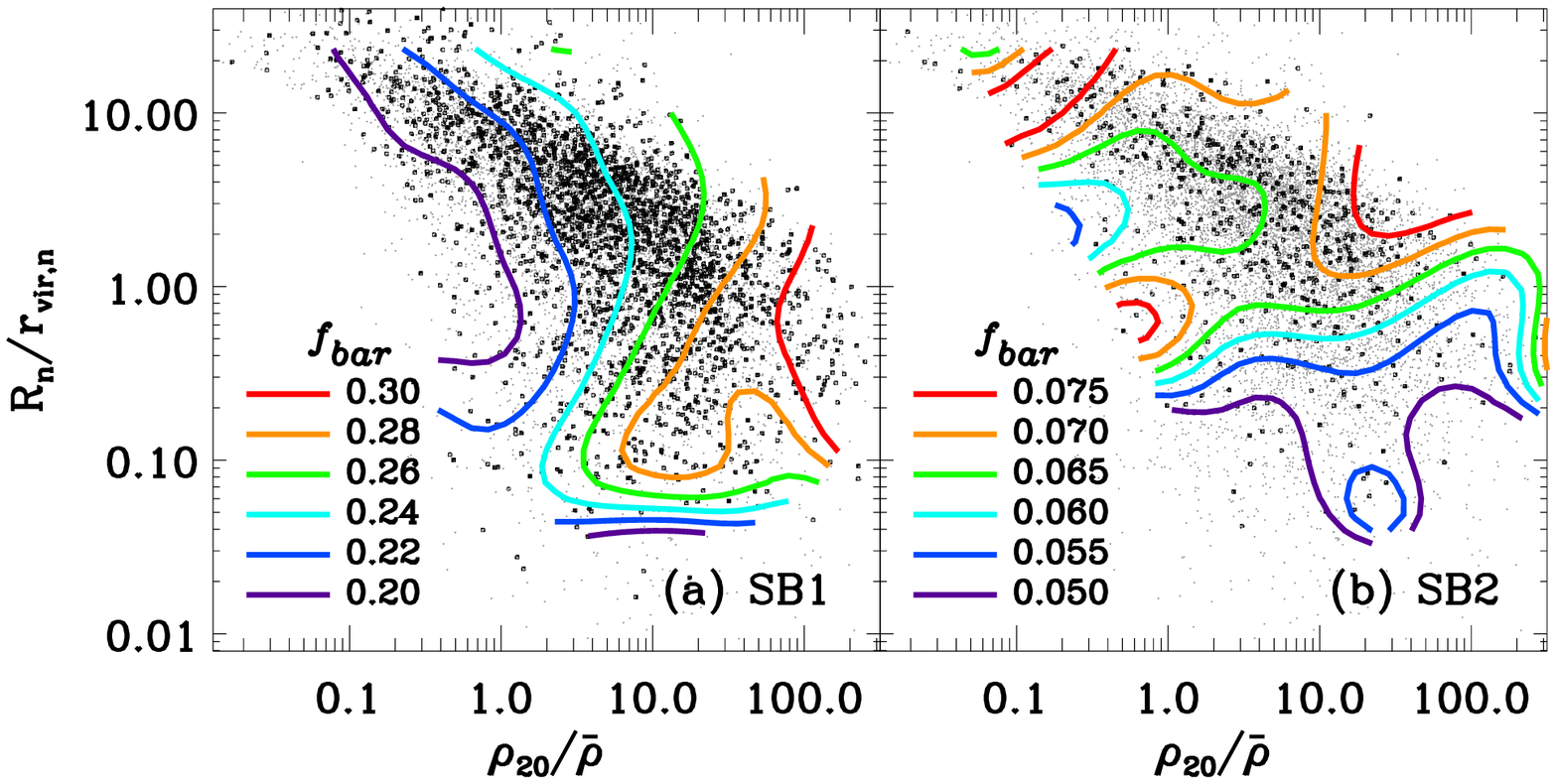}
\caption{Fraction of barred galaxies among face-on ($b/a>0.60$) late-type barred galaxies (contours) in two-dimensional
parameter spaces. Contours represent the fraction of SB1 galaxies (a) and those of SB2 galaxies (b).}
\label{locden_distnei}
\end{figure*}

\begin{figure}
\figurenum{13}
\centering
\includegraphics[scale=0.7]{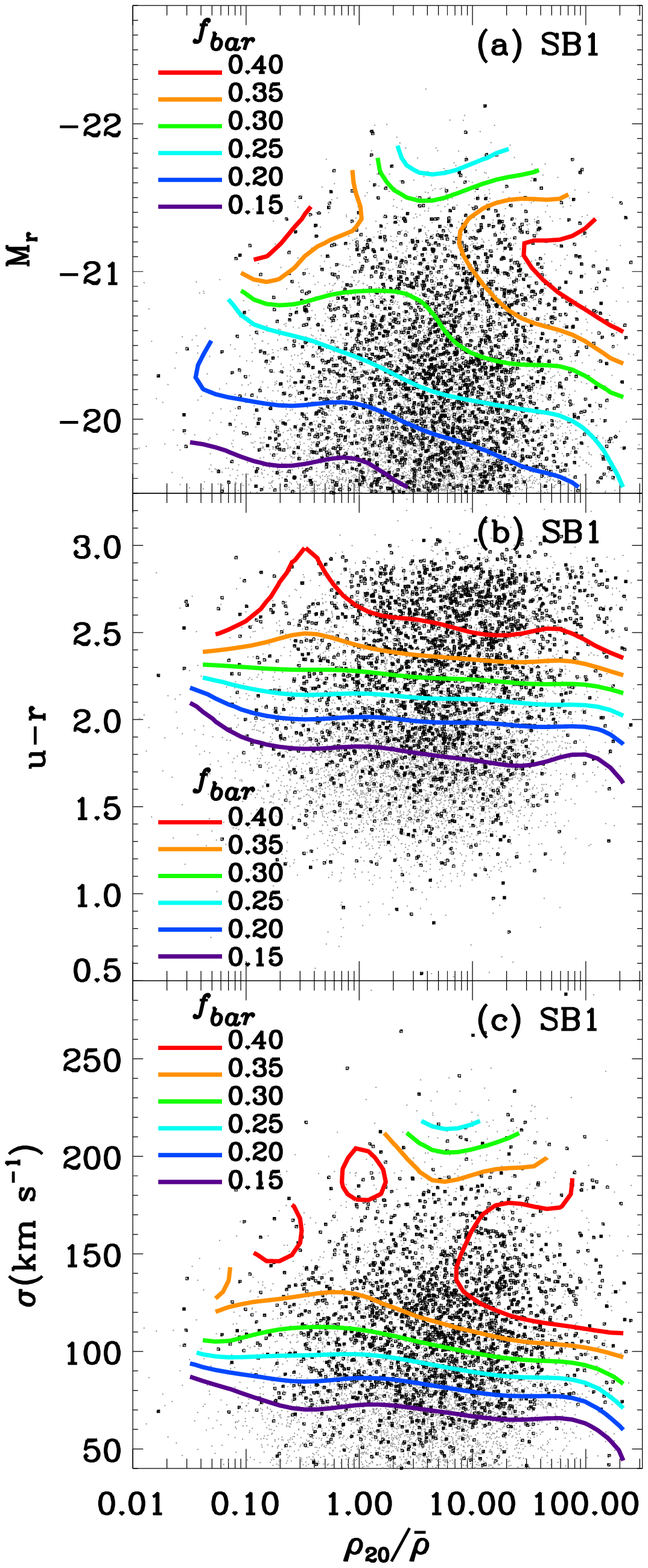}
\caption{Fraction of barred galaxies among face-on ($b/a>0.60$) late-type barred galaxies (contours) in two-dimensional
parameter spaces: (a) $\rtwenty/\brho$ versus $M_r$, (b) $\rtwenty/\brho$ versus $u-r$, and (c) $\rtwenty/\brho$ versus
$\sigma$. Black dots and grey dots represent, respectively, barred and non-barred late-type galaxies. Contours
represent constant barred galaxy fractions.}
\label{locden_params}
\end{figure}

\begin{figure*}
\figurenum{14}
\centering
\includegraphics[scale=0.65]{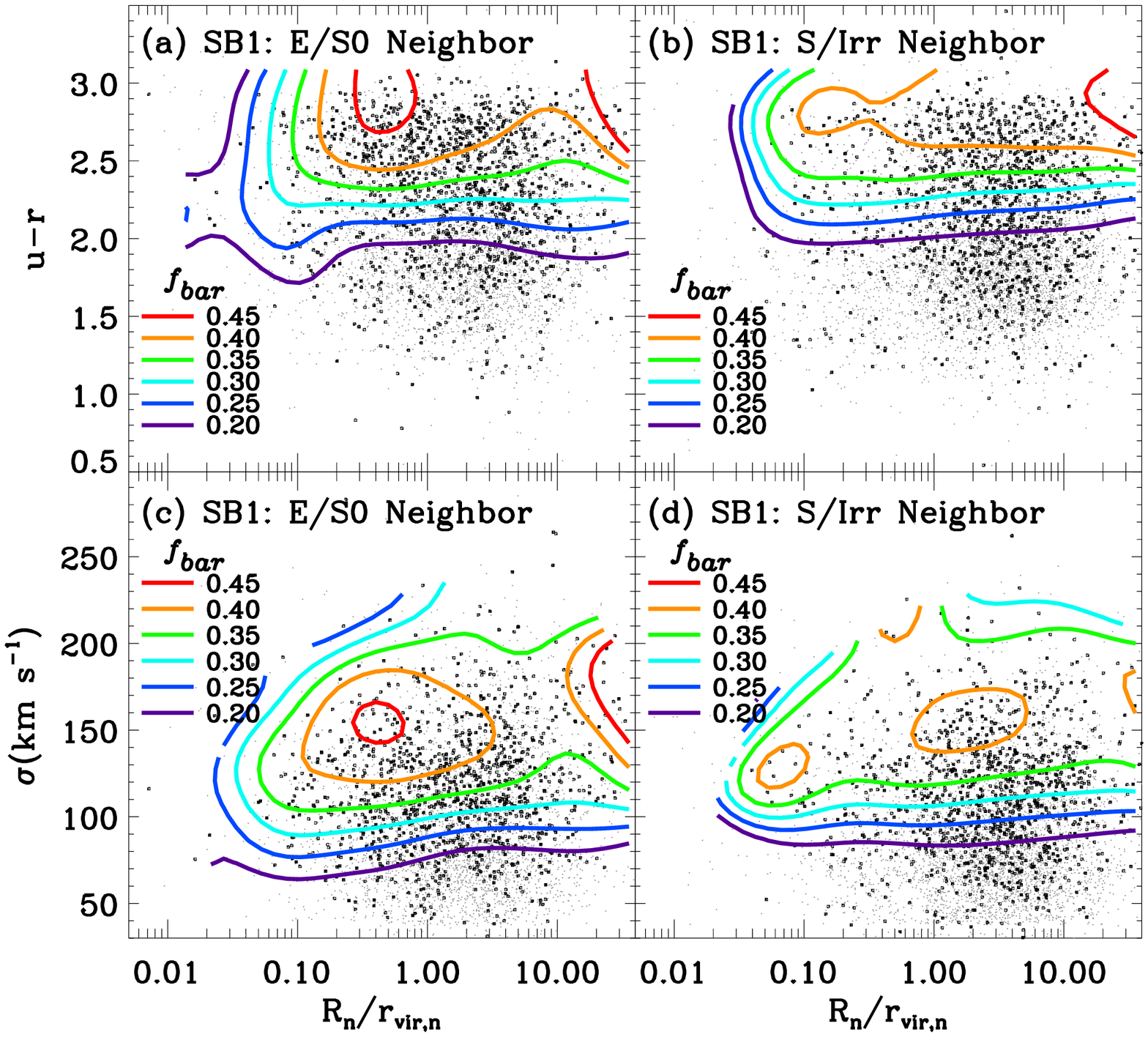}
\caption{Distribution of face-on ($b/a>0.60$) late-type galaxies in $\distnei$ versus the $u-r$ color
({\it upper}), the central velocity dispersion $\sigma$ ({\it lower}) spaces. {\it Left column}: galaxies
having an early-type neighbor, and {\it right column}: galaxies having a late-type neighbor.Black dots and
grey dots represent barred and non-barred late-type galaxies, respectively. Contours show constant SB1
galaxy fraction.}
\label{distnei_4params_sb1}
\end{figure*}

\begin{figure*}
\figurenum{15}
\centering
\includegraphics[scale=0.65]{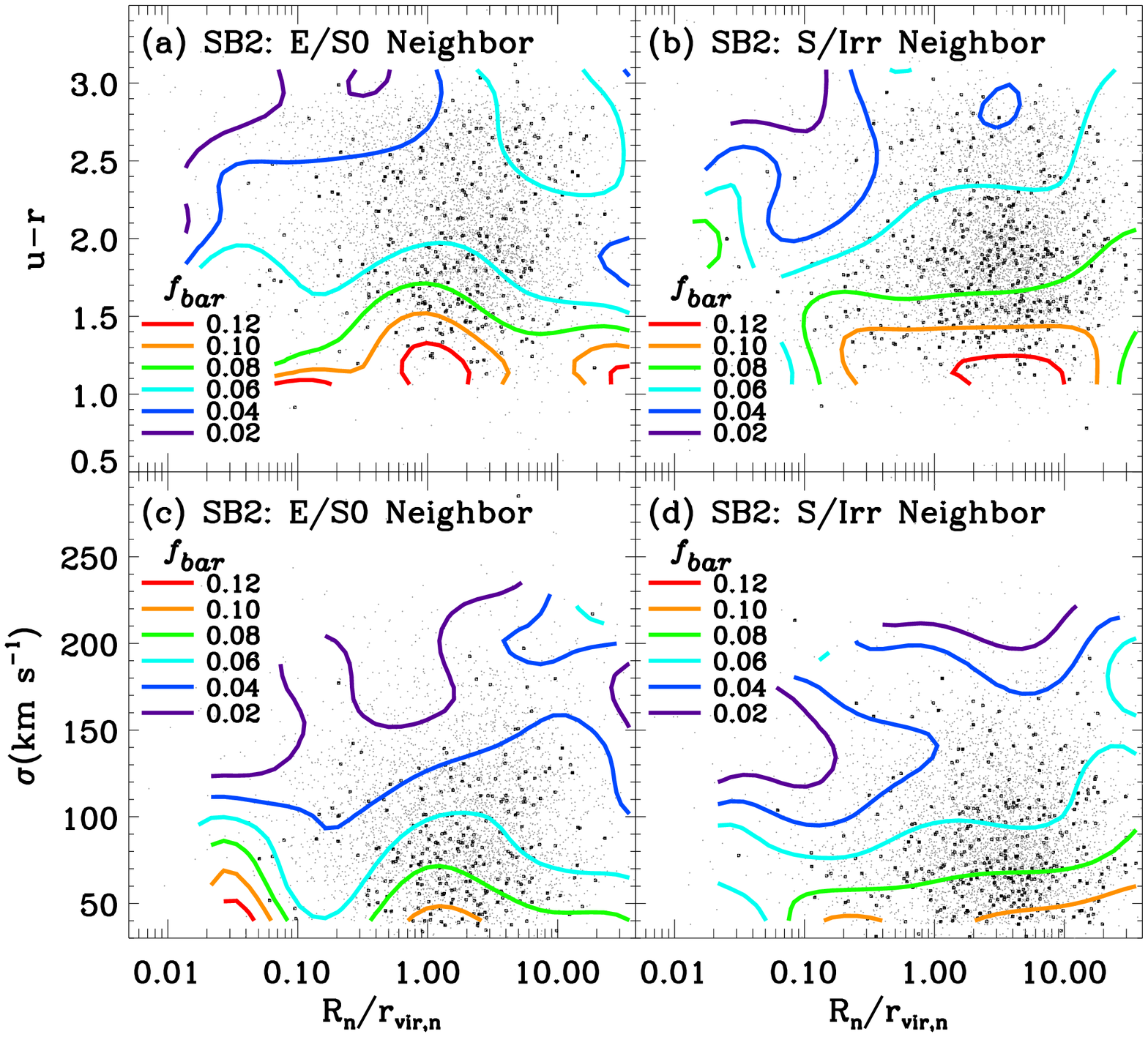}
\caption{The same as Figure \ref{distnei_4params_sb1} but for $\bfrsbw$.}
\label{distnei_4params_sb2}
\end{figure*}

\citet{thompson81} showed that $\bfr$ is significantly larger in the core of the Coma cluster than
in the outer part. However, recent studies suggested that there is no relation between $\bfr$
and local density \citep{mend08,agu2009,cheng09}.

We investigate the dependence of $\bfr$ on environment using a sample of 10,674 late-type galaxies ($b/a>0.60$)
that is much larger than those in previous studies, and measure $\bfr$ in different environments.
We choose three environmental parameters: large-scale background density ($\rtwenty/\brho$),
distance to the nearest neighbor galaxy normalized by the virial radius of the neighbor ($\distnei$),
and the morphology of the neighbor (an early-type or a late-type neighbor).

First, we divide our sample into three fixed absolute magnitude ranges:
$-19.5>M_{r}\ge-20.0$ mag, $-20.0>M_{r}\ge-20.5$ mag and $-20.5>M_{r}\ge-21.0$ mag. This is
to minimize the contamination by the correlation between luminosity and local density.
Figure \ref{envir_bf}a displays $\bfrsbo$ versus $\rtwenty/\brho$. It shows that $\bfrsbo$ appears
to be an increasing function of $\rtwenty/\brho$ in all three $M_r$ bins. This tendency is
more evident for $\rtwenty/\brho\ge1$. It seems that an environmental effect induces larger $\bfrsbo$
in higher density region. However, we will see below that this background density dependence
appears only through the correlation between galaxy properties and background density, and that
there is no evidence for direct correlation between $\bfrsbo$ and $\rtwenty/\brho$.

Figure \ref{envir_bf}b shows $\bfrsbo$ as a function of $\distnei$. At $R_{n}>0.2$ $r_{vir,n}$,
we find that $\bfrsbo$ decreases as the separation to the nearest neighbor galaxy increases.
Smaller $R_n$ corresponds statistically to higher $\rho_{20}$. Therefore, it is expected that
$\bfrsbo$ increases as $\distnei$ decreases, which similar to that $\bfrsbo$ increases as $\rtwenty/\brho$ increases.
However, when $\distnei<0.1$, $\bfrsbo$ drops significantly in all three $M_r$ bins.
This trend is more evident for the bright galaxies with $-20.5>M_r\ge-21.0$ mag.
This result indicates that strong bars can be destroyed or weakened by environmental effects
like tidal interactions with neighbors. On the other hand, there is no clear connection between the environmental
parameters and the fraction of SB2 galaxies as shown in Figure \ref{envir_bf}c and d.
We will discuss the relation between the bar fraction and environment in detail in Section 5.

To obtain a better understanding of the correlation between $\bfr$ and environmental parameters,
we inspect the behavior of $\bfr$ in several two-dimensional parameter spaces.
Figure \ref{locden_distnei}a shows the dependence of $\bfrsbo$ on two environmental parameters,
$\distnei$ and $\rtwenty/\brho$, simultaneously. Since the mean separation between galaxies decreases
as the background density increases, there is a statistical correlation between $R_n$ and $\rho_{20}$.
But in this figure one can investigate the dependence of $\bfr$ on one environmental parameter,
while other parameters are fixed. It is seen that $\bfrsbo$ depends dominantly on $\rtwenty/\brho$
except when the galaxy is very close to its nearest neighbor. When $\distnei<0.1$, contours become
horizontal, indicating that $\bfrsbo$ depends only on $\distnei$ not on $\rtwenty/\brho$. The critical distance
to the neighbor is about 0.1 $r_{vir,n}$, which is close to the merger scale \citep{park08}. Therefore,
this implies that strong bars are likely to be destroyed during strong tidal interactions.
In the case of SB2 galaxies, it seems that their fraction increases as $\distnei$ increases as shown in
Figure \ref{locden_distnei}b. However, this trend is not significant since the variation of $\bfrsbw$ across the whole
range of $\distnei$ is only $\sim0.02$. From this, we conclude that weak bars do not prefer a special environmental condition.

Also, in Figure \ref{locden_distnei}a, contours are nearly vertical when $\distnei>0.1$. The SB1 fraction
gradually increases as $\rtwenty/\brho$ increases, which can be interpreted that any environmental condition tends to
enhance the formation of a bar at the center of galaxies. However, as seen in Figure \ref{locden_params}, this
trend disappears when other parameters ($M_r$, $u-r$, and $\sigma$) are fixed. When $u-r$ color is fixed, $\bfrsbo$ is
nearly independent of $\rtwenty/\brho$ (see Figure \ref{locden_params}b). It demonstrates that $\rtwenty$-dependence of
$\bfrsbo$ in Figure \ref{locden_distnei}a is indeed caused by the fact that redder galaxies are likely to be located in
higher density regions \citep{ski08,bamford+09} and that $\bfrsbo$ has a positive correlation with $u-r$ color.
Also, when $\sigma$ is fixed, $\rtwenty$-dependence disappears as shown in Figure \ref{locden_params}c. While there exists
some residual dependence of $\bfrsbo$ on $\rtwenty$ when $M_r$ is fixed as shown in Figure \ref{locden_params}a.
It indicates that color and velocity dispersion are more influential parameters for occurrence of bars compared with $M_r$.

To extend our study on the $R_n$-dependence of $\bfr$, we adopt the third environmental parameter,
the morphology of the nearest neighbor. Galaxy properties are affected by both distance to the neighbor
and neighbor's morphology \citep{park08,pnh08,pnc09}; interactions with early-type neighbors
quench star formation activity of late-type galaxies, and late-type neighbors enhance the SFR of galaxies.
\citet{pnc09} found that $u-r$ color becomes redder and $\ha$ equivalent width decreases as late-type galaxies approach an early-type neighbor and that $u-r$ color becomes bluer and $W(\ha)$ becomes larger when late-type
galaxies approach a late-type neighbor galaxy. This bifurcation occurs at $R_n\simeq r_{vir,n}$
and it becomes particulary prominent at $R_{n}<0.1$ $r_{vir,n}$. This change in color and SFR depending on
neighbor's morphology must be caused by hydrodynamic interactions between the approaching galaxies.

Does this hydrodynamic interaction also affect the bar fraction? We inspect the behavior
of $\bfrsbo$ in $u-r$ versus $\distnei$ and $\sigma$ versus $\distnei$ spaces for galaxies with an early-type neighbor
and those with a late-type neighbor in Figure \ref{distnei_4params_sb1}. When $R_{n}>0.1$ $r_{vir,n}$, contours are nearly
horizontal, which means that $\bfrsbo$ has no dependence on distance to the neighbor galaxy. However,
when $R_{n}<0.1$ $r_{vir,n}$, contours becomes vertical to the $x$-axis, and $\bfrsbo$ decreases as $\distnei$ decreases.
We already found in Figure \ref{locden_distnei}a that $\bfrsbo$ significantly decreases when $\distnei<0.1$.
This tendency remains intact for both early- and late-type neighbor cases. When $R_{n}<0.1$ $r_{vir,n}$
and the neighbor is an early type, $\bfrsbo$ decreases as $\distnei$ decreases even though color becomes redder.
It seems that $\distnei$ is the most important parameter in determining $\bfrsbo$ when galaxies are undergoing strong
interactions with their neighbor galaxies. Since $\bfrsbo$ does not depend on neighbor's morphology,
hydrodynamic interaction between galaxies does not seem to affect the bar formation and evolution at the merger scales.

In Figure \ref{distnei_4params_sb2} we plot the same as Figure \ref{distnei_4params_sb1}, but for $\bfrsbw$.
It shows the strong dependence of $\bfrsbw$ on $u-r$ color and on $\sigma$. However, no obvious dependence on $\distnei$
is found even when $\distnei<0.1$, regardless of neighbor's morphology. In fact, there are not enough SB2 galaxies at $\distnei<0.1$ to investigate the neighbor's effect on $\bfrsbw$. When $R_{n}>0.1 r_{vir,n}$, we find no correlation
between $\distnei$ and $\bfrsbw$.

\section{Discussion}

In the previous sections, we investigated the dependence of $\bfrsbo$ and $\bfrsbw$ on physical parameters
of galaxies and environment. Among various physical parameters, $u-r$ color, central velocity dispersion, and
concentration index are the most important parameters affecting both $\bfrsbo$ and $\bfrsbw$.
The bar fraction of SB1 galaxies increases as $u-r$ color becomes redder while $\bfrsbw$ shows any opposite function
of $u-r$ color. On the other hand, $\bfrsbo$ reaches a peak at $\sigma\simeq 150\pm25 $ km s$^{-1}$, but $\bfrsbw$
becomes larger as $\sigma$ decreases. Also, $\bfrsbw$ increases as $c_{in}$ increases, while $\bfrsbo$ varies
depending on $c_{in}$ only for galaxies with $u-r>2.5$ or $\sigma>150$ km s$^{-1}$.

It seems that strong bars are preferentially located in the high density regions than in the low density regions.
However, we find that the large-scale background density does not directly affect $\bfrsbo$,
since $\rho_{20}$-dependence disappears when color or the central velocity dispersion is fixed (Figure
\ref{locden_params}). The nearest neighbor galaxy also hardly affects the variation of $\bfrsbo$
when a galaxy is located farther than 0.1 times virial radius of the neighbor galaxy. But, when
$R_{n}<0.1$ $r_{vir,n}$, $\bfrsbo$ abruptly drops as the separation between galaxies decreases,
and this trend appears regardless of neighbor morphology. On the other hand, in the case of weak bars we find
no evidence that $\bfrsbw$ and environmental parameters are correlated with each other.

\subsection{Implications for the secular evolution of barred galaxies}

\begin{figure}
\figurenum{16}
\centering
\includegraphics[scale=0.7]{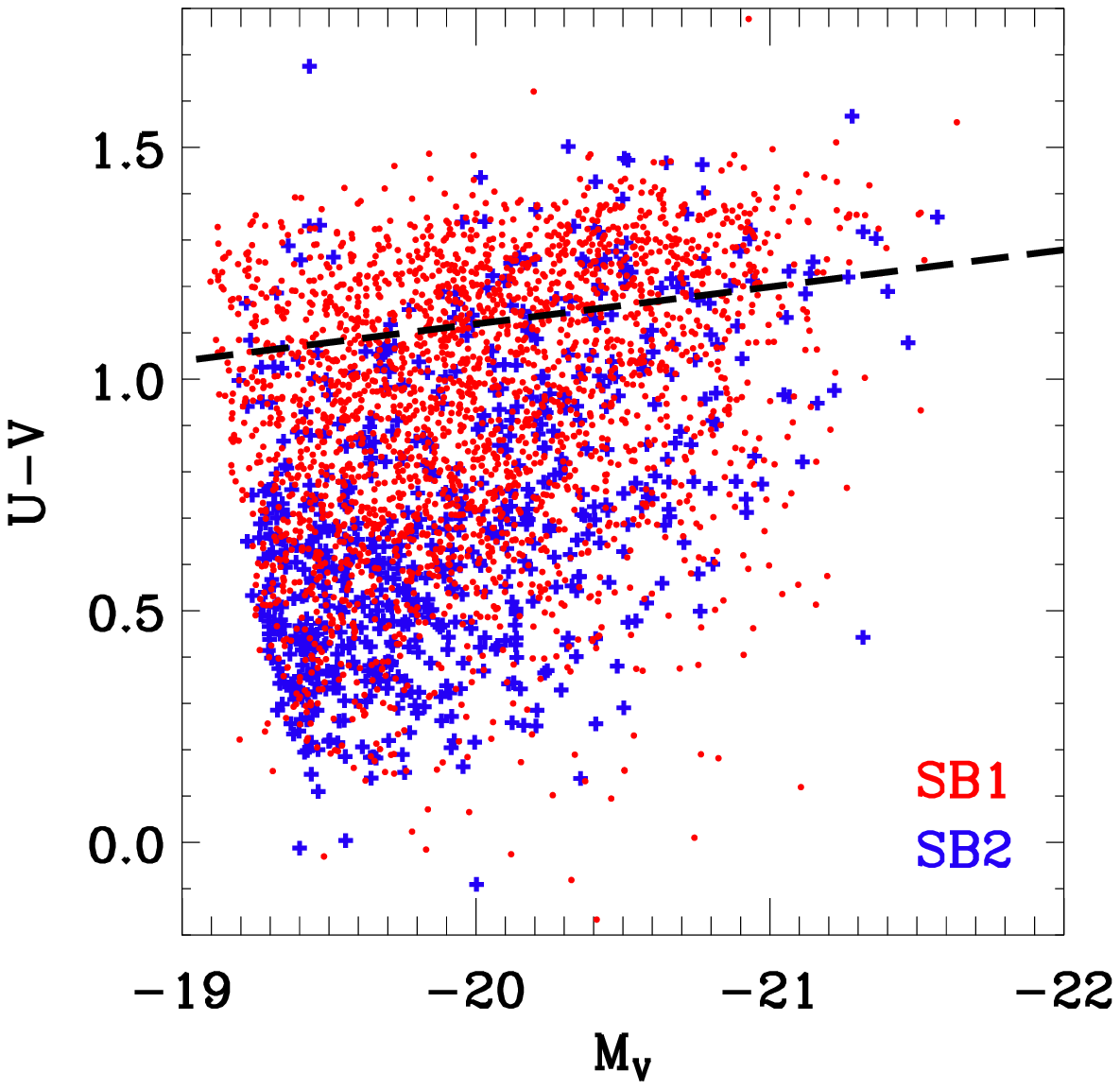}
\caption{Color-magnitude diagram for the late-type barred galaxies in our galaxy sample.
The dashed line represents $U-V=1.15-0.31z-0.08(M_{V}-5$log$h+20)$ \citep{bell04}.
Dots and crosses represent SB1 and SB2 galaxies, respectively.
Note that \citet{barazza08} used only the galaxies bluer than the dashed line,
while we used all galaxies in the diagram.}
\label{cmd_uvv}
\end{figure}

\begin{figure}
\figurenum{17}
\centering
\includegraphics[scale=0.55]{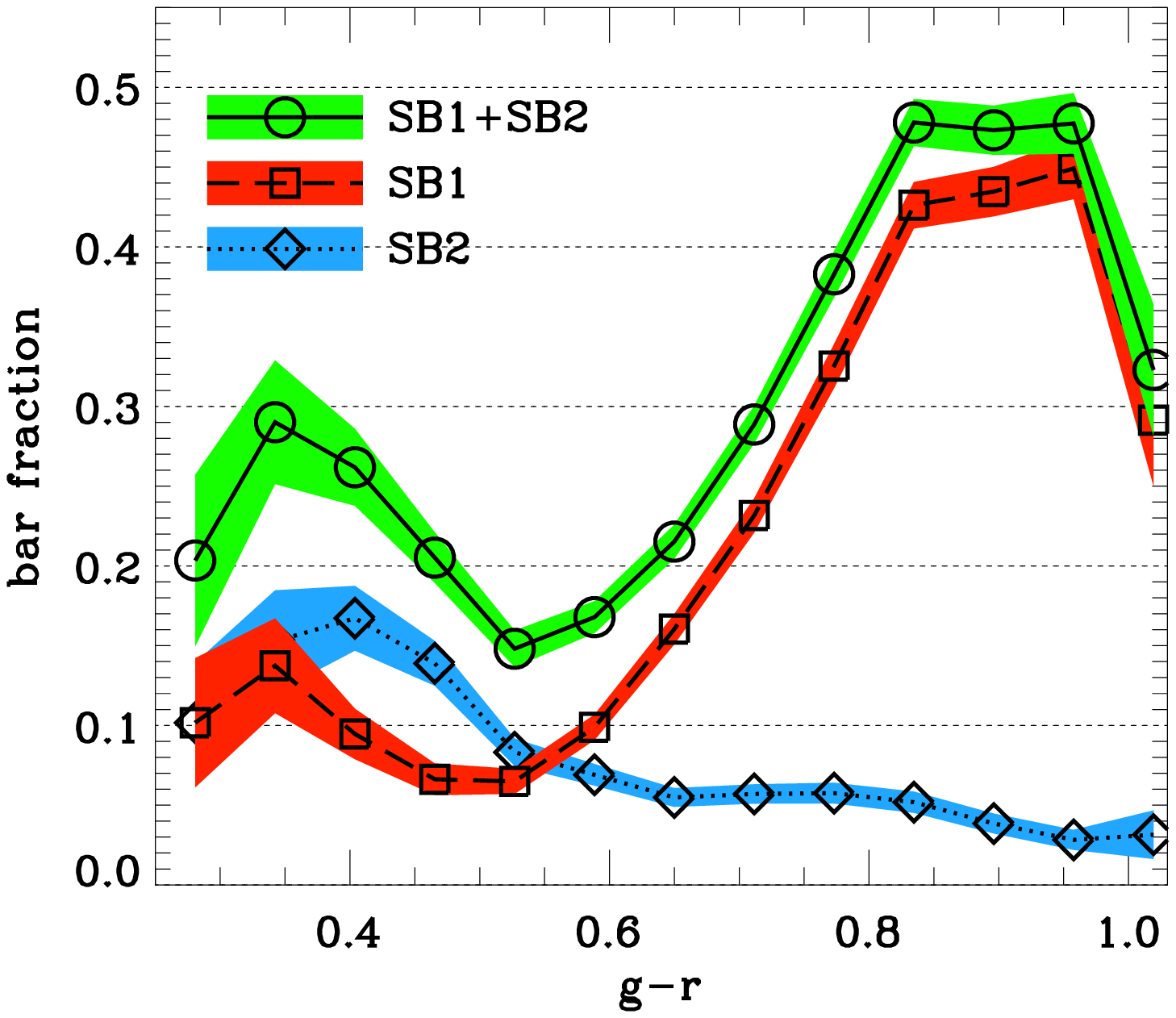}
\caption{The bar fraction as a function of $g-r$ color. Squares and diamonds represent the fraction of SB1 and SB2 galaxies,
respectively. While circles are the sum of two types of bar. The shade represents 1-$\sigma$ sampling error.}
\label{gr_bf}
\end{figure}

\subsubsection{Dependence of $\bfr$ on color}

One of our findings is that both $\bfrsbo$ and $\bfrsbw$ depend strongly on $u-r$ color and that $\bfrsbo$ is
higher in redder galaxies, while $\bfrsbw$ shows a peak in bluer galaxies. \citet{hoyle+11} found that galaxies having bars
(with their length $L<5$ h$^{-1}$ kpc) are bluer than those having a longer bar, which is consistent with our result.

However, our result is apparently opposed to the result given by \citet{barazza08} who showed that $\bfr$ increases
as $g-r$ color becomes bluer. The difference in galaxy sample selection may be a cause for this discrepancy.
\citet{barazza08} used a sample of 1,860 blue disk galaxies selected in the $U-V$ versus $M_V$ color-magnitude diagram.
Inevitably, red spirals are removed from their sample by the color cut. Figure \ref{cmd_uvv} shows the distribution
of late-type barred galaxies in our sample in the $U-V$ versus $M_V$ color-magnitude diagram. $U-V$ and $M_{V}$ are obtained
using transformation equations of \citet{jester+05}.
\citet{barazza08} used $U-V=1.15-0.31z-0.08(M_{V}-5$log$h+20)$ \citep{bell04}, represented by the dashed line in Figure
\ref{cmd_uvv}, to select disk galaxies. As shown in this figure, there are a significant number of red barred galaxies
above the dashed line. Therefore, their result is biased to bluer galaxies. Also, they studied the color dependence of $\bfr$
in the narrow color range of $0.2<g-r\lesssim0.7$ (see Figure 13d in \citealt{barazza08}).
The $g-r$ color of our late-type galaxies spans from $\sim$0.2 to $\sim1.0$ as shown in Figure \ref{gr_bf}.
This figure is very similar to Figure 3 in \citet{masters11}, but we distinguish weak bars from strong bars and
display the fraction of the two types. The excess of $\bfrsbw$ at blue end becomes more noticeable compared to that in Figure
\ref{params_bf}a. We find that $\bfrsbw$ decreases as $g-r$ increases, which is similar to the result in \citet{barazza08}.
Interestingly, at $g-r<0.5$, $\bfrsbo$ also shows a similar trend. Therefore, it becomes obvious that their color
dependence is only for blue disk galaxies. Our result shows a low peak of $\bfr$ (SB1+SB2) at $g-r\simeq0.35$, which is also
seen in Figure 1d of \citet{nair10b}.

From the fact that $\bfrsbo$ is higher in redder galaxies, we suggest two possibilities. First, bars could have an important role in
the formation of red late-type galaxies. \citet{masters11} also suggested an idea that transition from blue spirals to red spirals
may be due to turning off star formation by bars. This idea is supported by an argument that strong bars drive gas into central part
of galaxies more efficiently compared with weak bars \citep{del07}. In Figure \ref{ur_rabsmag_lineHa}b, we find that there are SB1
galaxies with higher SFR or with very low SFR. In terms of secular evolution, SB1 galaxies with very low SFR are considered to be
at the later stages where cold gas in disk had been used up by strong bars. Interestingly it shows a gap at $W(\ha)\simeq5$
${\rm \AA}$ where SB1 galaxies are relatively rare. This gap may suggest an observational evidence that the color transition
by a bar occurs quickly.

Second, red late-type galaxies are likely to provide better conditions for hosting a bar. In general
red late-type galaxies have earlier-type morphologies than blue late-type galaxies. We already checked that $\bfrsbo$ shows a
peak at small $c_{in}$, corresponding to early-type spirals. Some numerical simulations \citep{am02,ath2002,ath2003,ath2005}
suggested that bars in early-type spiral galaxies could be stronger and have a longer life than those in late-type spirals.
This expectation has a logical connection with our results that red late-type galaxies have a large $\bfrsbo$, that
blue late-type galaxies are likely to have a weak bar. Also, the excess of $\bfrsbo$ in red galaxies implies that
strong bars occur more frequently in the systems undergoing long passive evolution, which is also supported by the result
that $\bfrsbo$ drops when galaxies suffer from strong tidal interactions (as shown in Figure \ref{envir_bf}b,
\ref{locden_distnei}a and \ref{distnei_4params_sb1}).

\subsubsection{Dependence of $\bfr$ on central velocity dispersion}

\begin{figure}
\figurenum{18}
\centering
\includegraphics[scale=0.75]{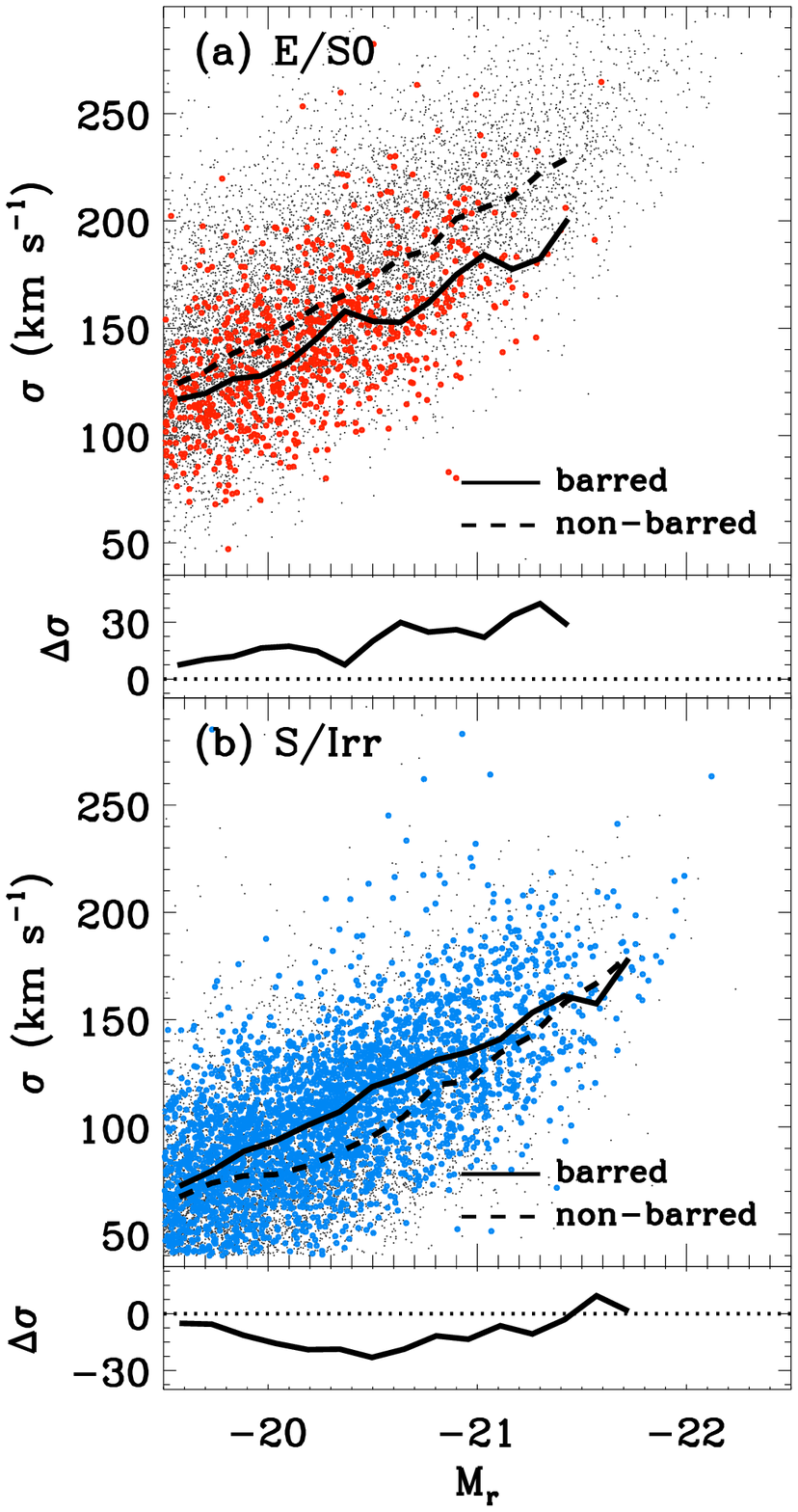}
\caption{Distribution of early-type (a) and late-type (b) galaxies in $M_r$ versus $\sigma$ spaces.
The median $\sigma$ depending on $M_r$ is drawn by solid lines for barred galaxies and
by dashed lines for non-barred galaxies. Barred galaxies in panel (a) indicate barred lenticulars (SB0s).
We also display the difference in $\sigma$ between barred and non-barred galaxies (non-barred$-$barred)
in the bottom panel.}
\label{rabsmag_vdisp}
\end{figure}

In galactic dynamics, the central velocity dispersion reflects the mass of a galaxy including dark
halo. The fact that $\bfrsbo$ reaches a maximum value at intermediate velocity dispersions shows
that occurrence of bars is most probable in intermediate mass systems. \citet{cameron10} also found
that $\bfr$ for the $0.2<z<0.6$ galaxies in the COSMOS field is higher in intermediate stellar mass
systems, which is consistent with our result. Some models suggested that bars are difficult to be generated
either if the disk is dynamically too hot, or if the disk mass is too small \citep{onp1973,ath2005,del07}.
This prediction is also consistent with our finding of the correlation between $\bfrsbo$ and $\sigma$.

Bar structures are supported by the long-axis stellar orbits rotating between Inner Lindblad Resonance (ILR)
and corotation radius \citep{combes85,knk2004}. Bars are thought to have an important role in driving gas toward
the central region of galaxies, and this phenomenon causes the central mass concentration (CMC) growth within ILR
\citep{norman96,saka99,snm99,shen2004,knk2004,ath2005,das08}. As bars evolve, the CMC
also grows, and this leads to increasing the velocity dispersion. Thus the velocity dispersion is higher in
early-type spirals than in late-type spirals \citep{gad05,gad06}. In addition, bars become longer
as they evolve \citep{ath2003,lau04,lau07,del07}. Therefore, it is reasonable that $\bfrsbo$ has an increasing function of
$\sigma$ for small $\sigma$ and that $\bfrsbw$ has a maximum at low $\sigma$.

On the other hand, as the CMC grows sufficiently, the ILR moves out to near the corotation
radius. Then the galaxy center becomes dynamically too hot and the bar-supporting long-axis orbit, so-called $x1$ orbit,
declines. Eventually the bars are disrupted \citep{ath2005,gad05,gad06}. Therefore, this effect could cause a decrease
of $\bfrsbo$ at high $\sigma$, which is consistent with our result that $\bfrsbo$ declines significantly at
$\sigma>175$ km s$^{-1}$.

In Figure \ref{rabsmag_vdisp} we plot $\sigma$ versus $M_r$ for early- and late-type galaxies.
This figure shows an observational evidence that bars disfavor the galaxies that are dynamically too hot.
Among late-type galaxies, barred galaxies generally have larger $\sigma$ than non-barred galaxies. However,
among early-type galaxies, barred lenticulars have much smaller $\sigma$ relative to non-barred early-type
galaxies at given $M_r$.
From this point of view, we find that bars could occur generally within a specific range of the central
velocity dispersion depending on $M_r$, and that bars are difficult to maintain for $\sigma$ above this range.
Therefore, our result supports the prediction that too large CMC causes the destruction of bars.

\subsubsection{Dependence of $\bfr$ on concentration index}

Relations between the bar fraction and concentration of galaxies are mentioned in several previous studies.
\citet{barazza08} and \citet{agu2009} found that the bar fraction is higher in less-concentrated systems than
more-concentrated systems. \citet{nair10b} also found a similar result only for Sbc and later type galaxies, but
for Sab and earlier type galaxies the bar fraction becomes higher in more-concentrated systems. They interpreted
their results that the bar destruction by a CMC works only for later type spirals with lower mass.

We find that $\bfrsbw$ is higher in less-concentrated systems (Figure \ref{params_bf}f, \ref{ur_conx_sigma}c,
and \ref{params_vdisp}f). This is consistent with results of previous studies. However, we find that in high mass
systems ($\sigma>150$ km s${}^{-1}$) $\bfrsbo$ also decreases as their host galaxies becomes more-concentrated systems,
but this trend is not found in lower mass systems. These result indicate that the bar destruction is associated with
not only the CMC but also some other factors, for example, the strength of bars and the mass of host galaxies.
Strong bars in relatively lower mass systems with $\sigma<150$ km s${}^{-1}$ can maintain their shape against an enhancement
of the CMC. However, the CMC is the crucial parameter for strong bars in high mass systems and for weak bars.

\subsection{Environmental dependence of bar evolution}

Several previous studies expected that bars are affected by environmental effect, such as tidal interactions
with companion galaxies. Thompson(1981) found that $\bfr$ is two times larger in the central part of the Coma cluster
than in the outer region, and argued that drastic tidal interactions in the core of Coma could generate bar instabilities
effectively. Later, some simulations succeeded in generating bars induced by galaxy interactions
\citep{noguchi1988,gca90,sundin1991,berentzen2004,elme05,agungonz09}. However, contrary to these expectations,
the enhancement of $\bfr$ in the high-density region was not found in recent observational studies
\citep{mend08,agu2009,cheng09,cameron10}. In particular, based on a sample of $\sim$ 800 bright galaxies ($M_{V}\le-18$)
in the Abell 901/2 supercluster, \citet{marinova09} measured the bar fraction in the optical band, and they did not find
any clear correlation between $\bfr$ and local density.

In this study we studied the dependence of $\bfr$ on environment using a galaxy sample much larger
than previous studies. The fraction of SB1 galaxies in the high-density regions with $\rtwenty/\brho\geq10$ is a factor of
$\sim$1.5 higher than in the low-density regions with $\rtwenty/\brho\leq0.1$. This gives an impression that
frequent tidal interactions in the high-density regions could form bars. But, it is found that the $\rho_{20}$-dependence
disappears almost entirely, if $u-r$ color or $\sigma$ is fixed, as shown in Figure \ref{locden_params}.
Also, in the case of weak bars, we do not find any clear dependence on $\rtwenty/\brho$.
This indicates that bars are neither stimulated nor destroyed in accordance with the background density,
and that the physical processes such as ram pressure heating, strangulation, and tidal interaction with groups and
clusters do not seem to be related to the bar phenomenon.

The environment set up by the nearest neighbor galaxy also does not influence $\bfr$ (both $\bfrsbo$ and $\bfrsbw$)
when the separation between galaxies is larger than 0.1 times the virial radius of the neighbor.
While strong bars seem to be destroyed by the strong tidal interactions with companions when a galaxy is located
within 0.1 times the virial radius (Figure \ref{distnei_4params_sb1}).
This result implies that it is difficult for bars to be maintained under strong tidal interactions. This is consistent
with an expectation of \citet{romanodiaz+08} who showed from numerical simulation that galaxy mergers weaken and shorten bars.
Strong tidal interactions can disturb stellar motion in the central part of galaxies, boost the central velocity dispersion,
and finally, destroy bars or weaken their strength. This interpretation is supported by \citet{cervantes10}'s finding that
the angular momentum of spiral galaxies, not only in its magnitude but also in its orientation, is affected by the presence
of a companion: the spin parameter decreases and a corresponding increase in alignment also appears once the galaxies are well within
the virial radius of their neighbor. In addition, the destruction of bars during strong tidal interactions occurs
regardless of neighbor's morphology, indicating that bars do not evolve through the hydrodynamic interactions
with nearby galaxies.
So far, some simulations expect that bar structures are generated by galaxy interactions \citep{noguchi1988,gca90,sundin1991,berentzen2004,elme05,agungonz09}.
However, we do not find any evidence for stimulating bars by environmental effects even in the case of weak bars.

\section{Summary}

We studied the dependence of bar fraction of late-type galaxies on internal physical properties of galaxies
and environmental factors. We used a volume-limited sample of galaxies, drawn from the SDSS DR7, brighter than
$M_{r}=-19.5$ mag and at redshift $0.02\le z\le 0.05489$.
We classify the galaxies in our sample into barred or non-barred ones by visual inspection.
To reduce contamination by internal extinction effects, we use only 10,674 late-type galaxies with axis ratio
$b/a>0.60$. Our major findings are as follows.

1. We find 3,240 barred galaxies ($\bfr=30.4\%$) that consist of 2,542 strong bars ($\bfrsbo=23.8\%$)
and 698 weak bars ($\bfrsbw=6.5\%$).

2. $\bfrsbo$ strongly depends on $u-r$ color: $\bfrsbo$ increases significantly as the color
becomes redder. Central velocity dispersion is another important parameter determining $\bfrsbo$.
The bar fraction of SB1 galaxies has a maximum value at intermediate velocity dispersions of
$\sigma_{max}=125\sim175$ km s$^{-1}$. These results suggest that strong bars are dominantly hosted
by intermediate-mass systems undergoing long secular evolution with little experience of recent interactions
or mergers.

3. $\bfrsbw$ also depends on $u-r$ and $\sigma$. But unlike SB1 galaxies, galaxies with bluer color or smaller $\sigma$
are more likely to host weak bars. These results mean that weak bars mainly inhabit later type spirals with low mass
and blue color.

4. $\bfrsbw$ is higher in less-concentrated systems than more-concentrated systems. $\bfrsbo$ also shows a similar
trend but only for high mass systems with $\sigma>150$ km s${}^{-1}$.

5. $\bfrsbo$ monotonically increases as the background density $\rtwenty/\brho$ increases. However, this
dependence disappears when $u-r$ or $\sigma$ is fixed. This indicates that the background density does not
play a direct role in the bar formation. In addition, we find that $\bfrsbw$ also do not depend on the background density.

6. Influence of the nearest neighbor galaxy on $\bfrsbo$ appears when the separation to the neighbor is smaller
than 0.1 times the virial radius of the neighbor. $\bfrsbo$ decreases as $\distnei$ decreases regardless of neighbor's
morphology. It indicates that it is difficult for galaxies to maintain strong bars during strong tidal interactions,
and that this phenomenon is of gravitational origin. The fraction of weak bars show little dependence on $\distnei$.

\vspace{1cm}

We thank H.S.Hwang for helpful comments and are grateful to J.H.Lee and J.B.Sohn for helping us
with the morphology classification.
C.B.P. acknowledges the support of the National Research Foundation of Korea (NRF)
grant funded by the Korea government (MEST) (No. 2009-0062868).
M.G.L. was supported in part by Mid-career Research Program through NRF grant funded by the MEST (No.2010-0013875).
YYC wss supported by the National Research Foundation of Korea to the Center for Galaxy
Evolution Research.
Funding for the SDSS and SDSS-II has been provided by the Alfred P. Sloan Foundation,
the Participating Institutions, the National Science Foundation, the U.S. Department of Energy,
the National Aeronautics and Space Administration, the Japanese Monbukagakusho,
the Max Planck Society, and the Higher Education Funding Council for England.
The SDSS Web site is http://www.sdss.org/.
The SDSS is managed by the Astrophysical Research Consortium for the Participating Institutions.
The Participating Institutions are the American Museum of Natural History, Astrophysical Institute Potsdam,
University of Basel, Cambridge University, Case Western Reserve University, University of Chicago,
Drexel University, Fermilab, the Institute for Advanced Study, the Japan Participation Group,
Johns Hopkins University, the Joint Institute for Nuclear Astrophysics, the Kavli Institute for Particle
Astrophysics and Cosmology, the Korean Scientist Group, the Chinese Academy of Sciences (LAMOST),
Los Alamos National Laboratory, the Max-Planck-Institute for Astronomy (MPIA), the Max-Planck-Institute for Astrophysics (MPA),
New Mexico State University, Ohio State University, University of Pittsburgh, University of Portsmouth,
Princeton University, the United States Naval Observatory, and the University of Washington.

\end{document}